\documentclass[iop,apj]{emulateapj}
\usepackage{booktabs}

\usepackage{graphicx,epstopdf,hyperref,color}

\newcommand{\Ion}[2]{#1{\,\scriptsize #2}}
\newcommand{\Teff}{\mbox{$T_{\mathrm{eff}}$}}

\newcommand{\Rfuv}{\mbox{$R'_{\mathrm{FUV}}$}}
\newcommand{\Rnuv}{\mbox{$R'_{\mathrm{NUV}}$}}
\newcommand{\Efuv}{\mbox{$E_{\mathrm{FUV}}$}}
\newcommand{\Enuv}{\mbox{$E_{\mathrm{NUV}}$}}

\begin{document}
\title{The UV Emission of Stars in LAMOST Survey I. Catalogs}

\author{Yu Bai\altaffilmark{1}}
\author{JiFeng Liu\altaffilmark{1,2}}
\author{James Wicker\altaffilmark{3}}
\author{Song Wang\altaffilmark{1}}
\author{JinCheng Guo\altaffilmark{4,1}}
\author{YuXiang Qin\altaffilmark{5}}
\author{Lin He\altaffilmark{1,2}}
\author{JianLing Wang\altaffilmark{1}}
\author{Yue Wu\altaffilmark{1}}
\author{YiQiao Dong\altaffilmark{1}}
\author{Yong Zhang\altaffilmark{6}}
\author{Yonghui Hou\altaffilmark{6}}
\author{Yuefei Wang\altaffilmark{6}}
\author{Zihuang Cao\altaffilmark{1}}

\altaffiltext{1}{Key Laboratory of Optical Astronomy, National Astronomical Observatories, Chinese Academy of Sciences,
       20A Datun Road, Chaoyang Distict, Beijing 100012, China; ybai@nao.cas.cn}
\altaffiltext{2}{College of Astronomy and Space Sciences, University of Chinese Academy of Sciences, Beijing 100049, China}
\altaffiltext{3}{National Astronomical Observatories, Chinese Academy of Sciences, 20A Datun Road, Chaoyang Distict, Beijing 100012}
\altaffiltext{4}{Department of Astronomy, Peking University, Beijing 100871, China}
\altaffiltext{5}{School of Physics, Parkville, Victoria 3010, Australia}
\altaffiltext{6}{Nanjing Institute of Astronomical Optics \& Technology, National Astronomical Observatories, Chinese Academy of Sciences, Nanjing 210042, China}

\begin{abstract}
We present the ultraviolet magnitudes for over three million stars in the LAMOST survey, 
in which 2,202,116 stars are detected by $GALEX$. For 889,235
undetected stars, we develop a method to estimate their
upper limit magnitudes. The distribution of (FUV $-$ NUV)
shows that the color declines with increasing effective temperature
for stars hotter than 7000 K in our sample, while the trend
disappears for the cooler stars due to upper atmosphere
emission from the regions higher than their photospheres. For stars
with valid stellar parameters, we calculate the UV excesses with
synthetic model spectra, and find that the (FUV $-$ NUV) vs. {\Rfuv}
can be fitted with a linear relation and late-type dwarfs tend to
have high UV excesses. There are 87,178 and 1,498,103 stars detected
more than once in the visit exposures of $GALEX$ in the FUV and NUV,
respectively. We make use of the quantified photometric errors to
determine statistical properties of the UV variation, including
intrinsic variability and the structure function on the timescale of
days. The overall occurrence of possible false positives is below
1.3\% in our sample. UV absolute magnitudes are calculated for stars
with valid parallaxes, which could serve as a possible reference
frame in the NUV. We conclude that the colors related to UV provide
good criteria to distinguish between M giants and M dwarfs, and the
variability of RR Lyrae stars in our sample is stronger than that of
other A and F stars.
\end{abstract}


\section{Introduction}


The Morgan-Keenan (MK) spectral classification system has
served as the primary system for classifying stars. The
classification is based on spectral continuum and features from
the convection or radiative diffusion of a heated
photosphere \citep{Morgan43}. The higher regions of the stellar
atmosphere (chromosphere, transition region, and corona)
are often dominated by more violent non-thermal physical process
mainly powered by the magnetic field. These
processes related to the magnetic field could result in
observable flares and starspots, and lead to discrepancy
between observations and the general reference frame defined by the
MK classification.

Since stellar activity is a complex phenomenon, it has many
proxies, e.g., \Ion{Ca}{II} H\&K lines or H$_\alpha$
equivalent width. Compared to spectral proxies, photometric
indicators have the ability to cover much more area of the
sky, and further provide a more efficient diagnosis of
stellar activity \citep{Olmedo15}. Because the UV waveband
is particularly sensitive to hot plasma emission ($\sim$10$^4-10^6$
K), the UV domain is ideal for investigating
stellar activity. The availability of UV photometry as an
activity indicator has been explored for Sun-like stars
\citep{Findeisen11} and M dwarfs \citep{Jones16}. Most of
stellar activity is time dependent in the range from minutes to
days \citep{Kowalski09,Wheatley12}. Repeated observations
in the UV band provide us with an opportunity to
characterize the time dependence of stellar activity, which
is still poorly understood due to the previous small size
of samples.



\begin{figure}
   \centering
   \includegraphics[width=0.45\textwidth]{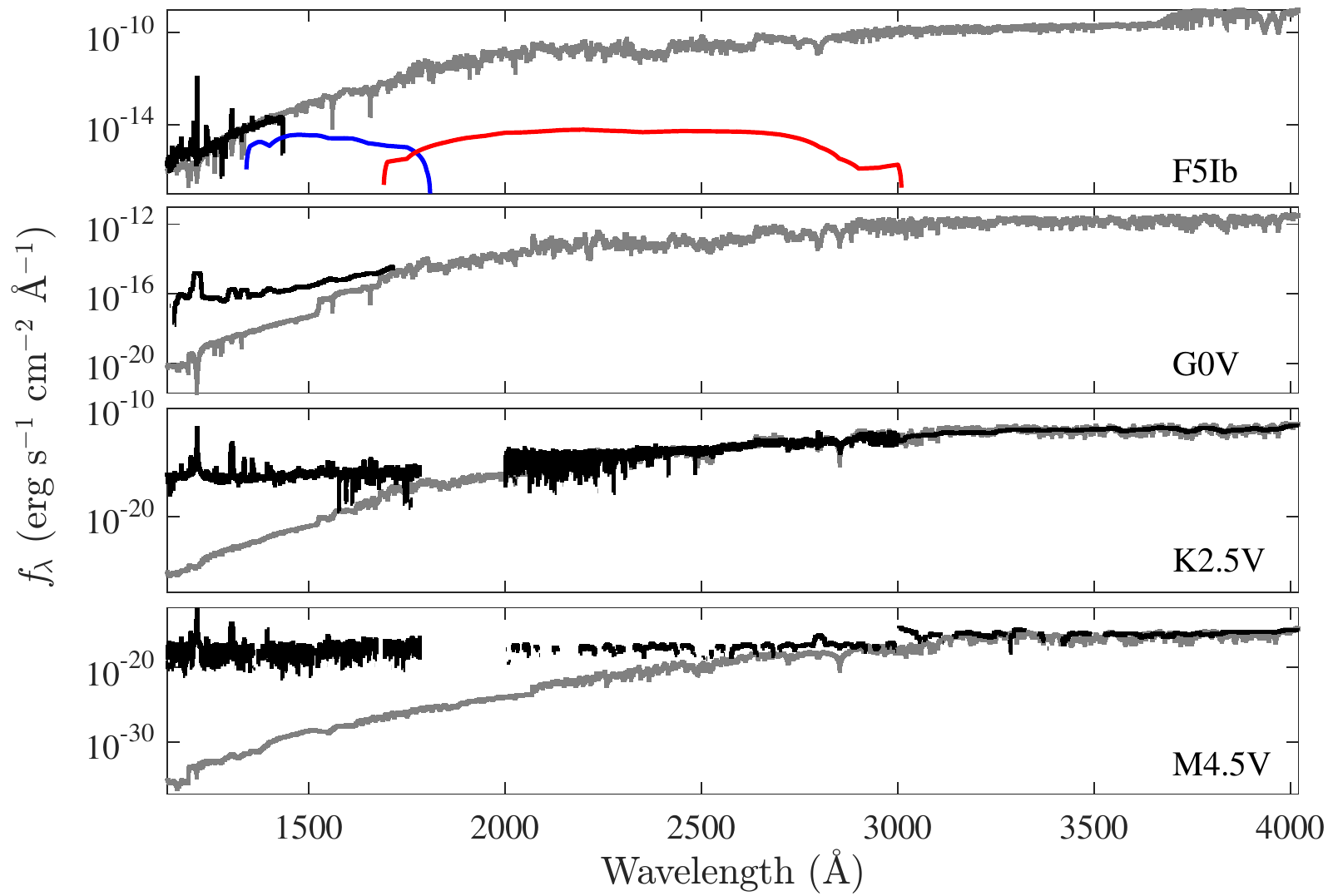}
   \caption{Examples of the UV emission that exceeds the photospheric models. The black lines are spectra from the $HST$
            Archive Data \footnote{https://archive.stsci.edu/hst/}, and the gray lines are PHOENIX synthetic models.
            The synthetic spectra are converted to observed flux with the $V$ band magnitudes.
            First panel: $\alpha$ Per, $HST$ proposal 14349 by Thomas Ayres. The FUV and NUV filters of $GALEX$
            are plotted in blue and red respectively.
            The star has strong emission lines in the FUV that are unpredicted by the model.
            It may have a white dwarf companion, but there is still no solid evidence \citep{Ayres17}.
            Second panel: HD 209458, $HST$ proposal 10081 by Alfred Vidal-Madjar.
            Third and forth panels: HD 40307 and GJ 1214. The spectra are extracted from MUSCLES \citep{Loyd16}
            \footnote{https://archive.stsci.edu/prepds/muscles/}. We delete the points with flux below zero.
   \label{HST}}
\end{figure}

Nevertheless, data from the $Galaxy$ $Evolution$ $Explorer$
($GALEX$; \citealt{Martin05,Morrissey07}) enable us to constrain
stellar UV behavior of a large sample in both near UV (NUV;
1771$-$2831 \AA) and far UV (FUV; 1344$-$1768 \AA) that cover the
spectral activity indicators of \Ion{C}{II}, \Ion{C}{IV},
\Ion{Mg}{II}, and \Ion{Fe}{II}. We present examples in Figure
\ref{HST} in order to shown the inadequacy of photospheric models to
explain the observed flux by the $GALEX$. The emission from F stars
begins to exceed the flux predicted by the photospheric model
(the first panel in Figure \ref{HST}).
The spectrum of GJ 1214 is shown in the fourth panel, in
which the UV excess is strongest. This result is similar to Figure
12 in \citet{Loyd16} that previously showed departures of the
observed UV spectrum of GJ 832 from the PHOENIX model spectrum. The
$GALEX$ survey provides us with an efficient path to
investigate the discrepancy between observations and models
for a large number of stars. $GALEX$ also presents a unique
opportunity to study stellar time-domain characteristics in the UV
\citep{Gezari13}, e.g., M-dwarf flare stars \citep{Welsh07} and RR
Lyrae stars \citep{Welsh05,Wheatley08}.

In this paper, we take advantage of the $GALEX$ archive data to
study stellar UV emission and variation. The sample is extracted
from the survey of the Large Sky Area Multi-Object Fiber
Spectroscopic Telescope (LAMOST, \citealt{Cui12}), which mainly aims
at understanding the structure of the Milky Way \citep{Deng12}.
The survey has obtained more than four million spectra of
stars from October 2011 to May 2015 (Section~\ref{data}). The
spectral types of these stars and accurate estimation of the stellar
parameters are produced by the LAMOST standard data processing
pipeline \citep{Luo15}.

Extinction is crucial in our study,  since it is about ten time
higher in the UV than in the near-infrared (IR) band \citep{Yuan13}.
In Section \ref{data}, we adopt the extinction measured with
a Bayesian method \citep{Wang16a}, and the Rayleigh-Jeans Color
Excess (RJCE) method \citep{Majewski11,Cutri13}. The infrared
counterparts of our sample are cross identified with the
catalog associated with the Wide-field Infrared Survey
Explorer ($WISE$).

For stars observed but undetected by $GALEX$, we develop a method to
estimate their upper limit magnitudes in the FUV and NUV in
Section~\ref{data}. In Section~\ref{Content}, we present the catalog
and the parameters to describe UV excess and variability. We then
analyze the contamination of other UV emitters and present
our methods to estimate false positives. Section~\ref{app} presents
the discussion and some scientific applications of our catalog. The
summary and further work are given in Section~\ref{sum}.
Throughout this paper, we adopt the extinction coefficients provided
by \citet{Yuan13} to correct Galactic extinction.

\section{Data} \label{data}


\subsection{LAMOST}
The design of LAMOST enables it to take 4000 spectra in a single
exposure to a limiting magnitude as faint as $r$ $=$ 19 at the
resolution R = 1800. In 2015.05.30, LAMOST finished its
third year survey (the third data release; DR3) and the all-sky
coverage is $\sim$ 35\%. DR3 of the LAMOST general catalog contains
objects from the LAMOST pilot survey and the three-year regular
survey, in all 5,755,216 objects included that are
observed multiple times. Objects with unique $designation$
in the catalog are selected as unique objects. The stars of our
sample are extracted from the catalog with the criteria of $class$ =
STAR and $subclass$ = from O to M \citep{Luo15},
which yields 4,010,635 stars. 

The stellar parameters of \Teff, log $g$, and [Fe/H] are extracted
from the A, F, G and K type star catalog, which was produced by the
LAMOST stellar parameter pipeline (LASP, \citealt{Wu14}).
The molecular indices are selected from the M star catalog.

\subsection{$WISE$}
We cross match the stars with the $ALLWISE$ catalogue
\citep{Cutri13} in order to estimate the extinction with the RJCE
method. We apply the following criteria to retrieve photometry in
the bands of $W2$ and $H$: the match radius is set to 5\arcsec; the
extended source flat (ext\_flg) is set to 0; the contamination and
confusion flags (cc\_flags) are set to 0; and the photometric
quality flag (ph\_qual) is set to A \citep{Wang16a}. Every star in
our sample has an infrared counterpart in the $ALLWISE$ catalogue,
but 23,388 stars do not have magnitudes in the $H$ band. We reject
these stars, and there are 3,987,247 stars left.

\subsection{$GALEX$}

We cross match the stars to the $GALEX$ Release 6 and 7 (GR6/GR7) in
order to obtain their photometry in the FUV and NUV. With
the help of the Catalog Archive Server Jobs
(CasJobs)\footnote{http://galex.stsci.edu/casjobs/}, nearby matching
is made to the table of $PhotoObjAll$ using the following criteria:
the match radius $\leq$ 5\arcsec \citep{Gezari13}; nuv\_artifact and
fuv\_artifact $\leq$ 1; the distance from the center of view $\leq$
0\arcdeg.55 \citep{Jones16}; and signal-to-noise ratio of the
magnitude $\geq$ 2, which yields 1,805,254 stars detected in the
co-added exposures.

In order to obtain magnitudes from the visit exposures, the
nearby matching is made to the table of $VisitPhotoObjAll$ with the
match radius $\leq$ {5\arcsec} and the distance from the center of
view $\leq$ {0\arcdeg.55}. The criterion of uv\_artifact is not used
here, since we want to select as many time-domain
observations as possible. We flag the detections with
uv\_artifact $>$ 1 in our catalog. The result includes 2,202,116
stars detected in the co-added and/or the visit exposures. The
nearest counterpart is selected for the stars with more than one
counterpart within {5\arcsec} in the same exposure, and they are
marked in the catalog.
There are 87,178 and 1,498,103 stars detected more than once in the visit exposures in the
FUV and NUV, respectively.

\begin{figure*}
   \centering
   \includegraphics[width=0.9\textwidth]{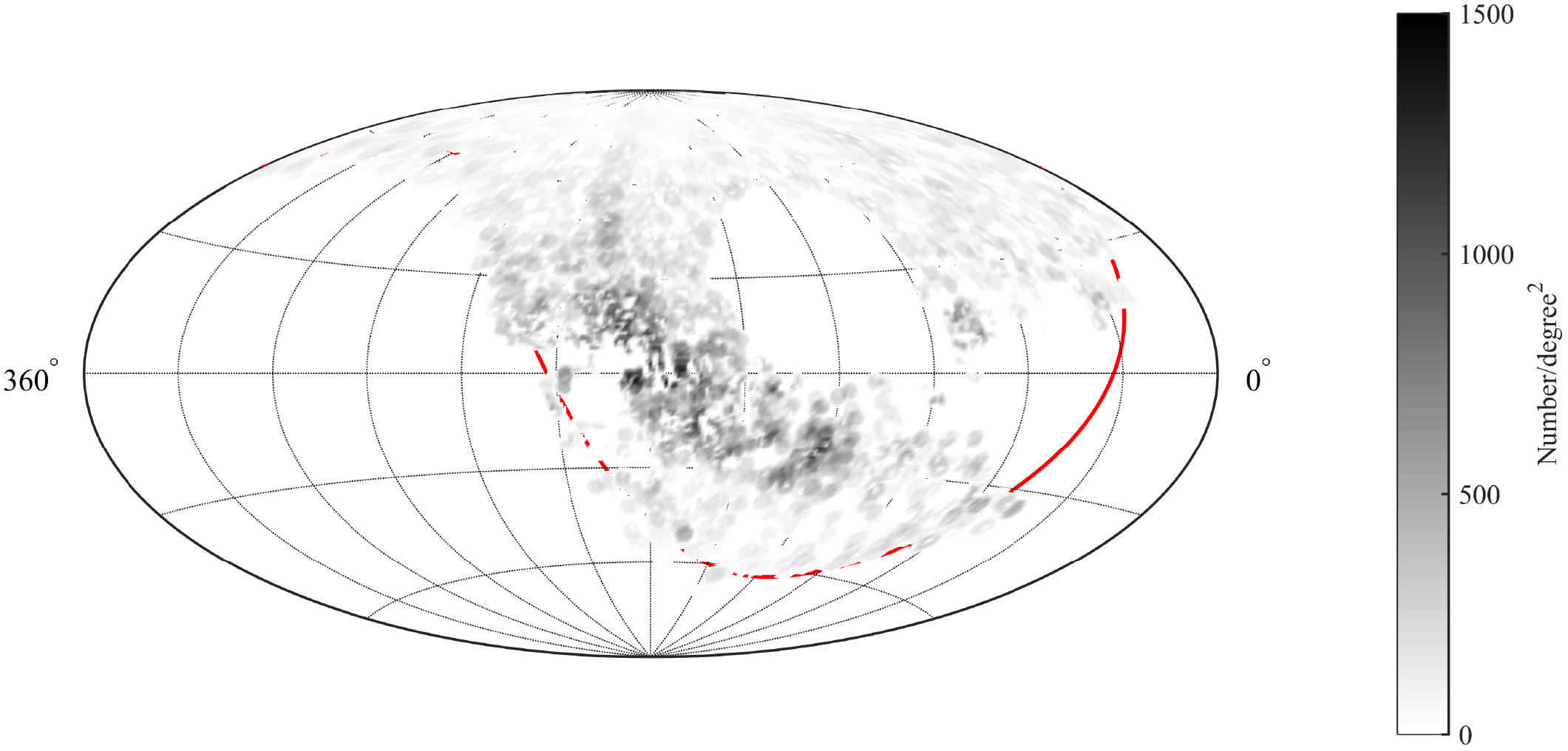}
   \caption{ The density map of stars in our catalog in Galactic coordinate.
   The celestial equator is shown as the red line.
   \label{gc_Lamost_ALLWISE_GALEX}
    }
\end{figure*}

The observed but undetected stars are extracted from the
tables of $PhotoExtract$ and $VisitPhotoExtract$, with the match
radius {$\leq$ 5\arcsec} and the distance from the center of view
{$\leq$ 0\arcdeg.55}. This results in an
additional 889,235 stars. The density map of all the 3,091,351
stars is shown in Figure~\ref{gc_Lamost_ALLWISE_GALEX}.

\subsection{Upper Limit}
\citet{Gezari13} defined the upper limit magnitude
determined as a function of exposure times, and the brightness of
the sky background. In their calculation, the background is
a constant. However, the UV background emission, which is dominated
by zodiacal light and diffuse galactic light, significantly depends
on Galactic position \citep{Murthy14}.

In order to calculate the position and exposure-time dependent
background, we extract sources from the table of $VisitPhotoObjAll$
with the following criteria: distance from the center of view $\leq$
0\arcdeg.55; signal-to-noise ratios (S/N) $\lesssim$ 1.8 and
$\lesssim$ 1.2 in the FUV and NUV. This method is similar to that
used in \citet{Ansdell15}.

\begin{figure*}
   \centering
   \includegraphics[width=0.9\textwidth]{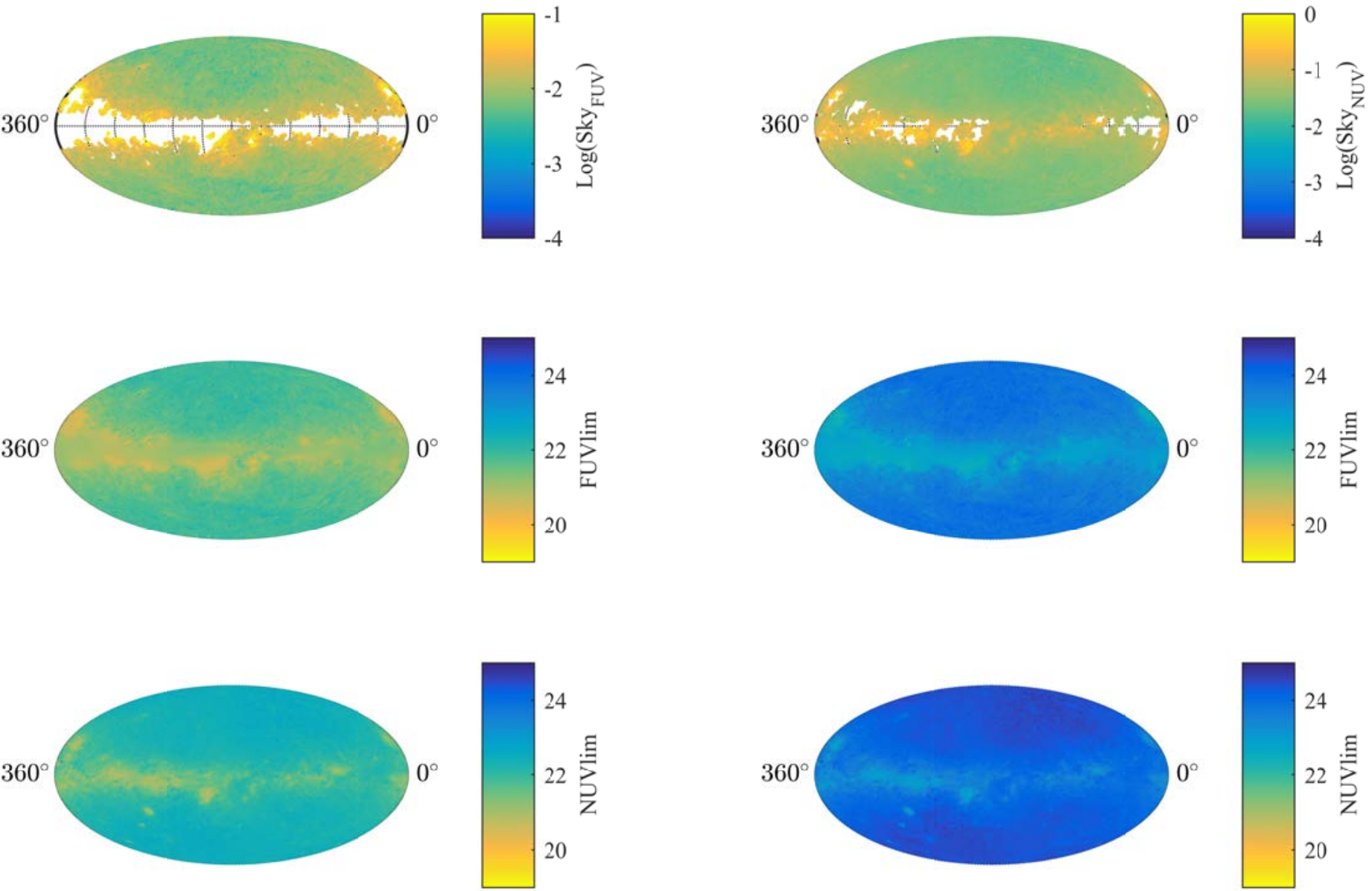}
   \caption{ The UV background and upper limit magnitudes in Galactic coordinates.
             Upper panels: aitoff projection of the Galactic coordinate of the fitted sky background (cts s$^{-1}$).
             Middle panels: the interpolated upper limit magnitudes in FUV with the exposure time
             from 30 to 50 sec and from 1680 to 1710 sec.
             Lower panels: the same as the Middle panels in NUV.
   \label{gc_Up_FUVNUV}
    }
\end{figure*}

\begin{figure}
   \centering
   \includegraphics[width=0.45\textwidth]{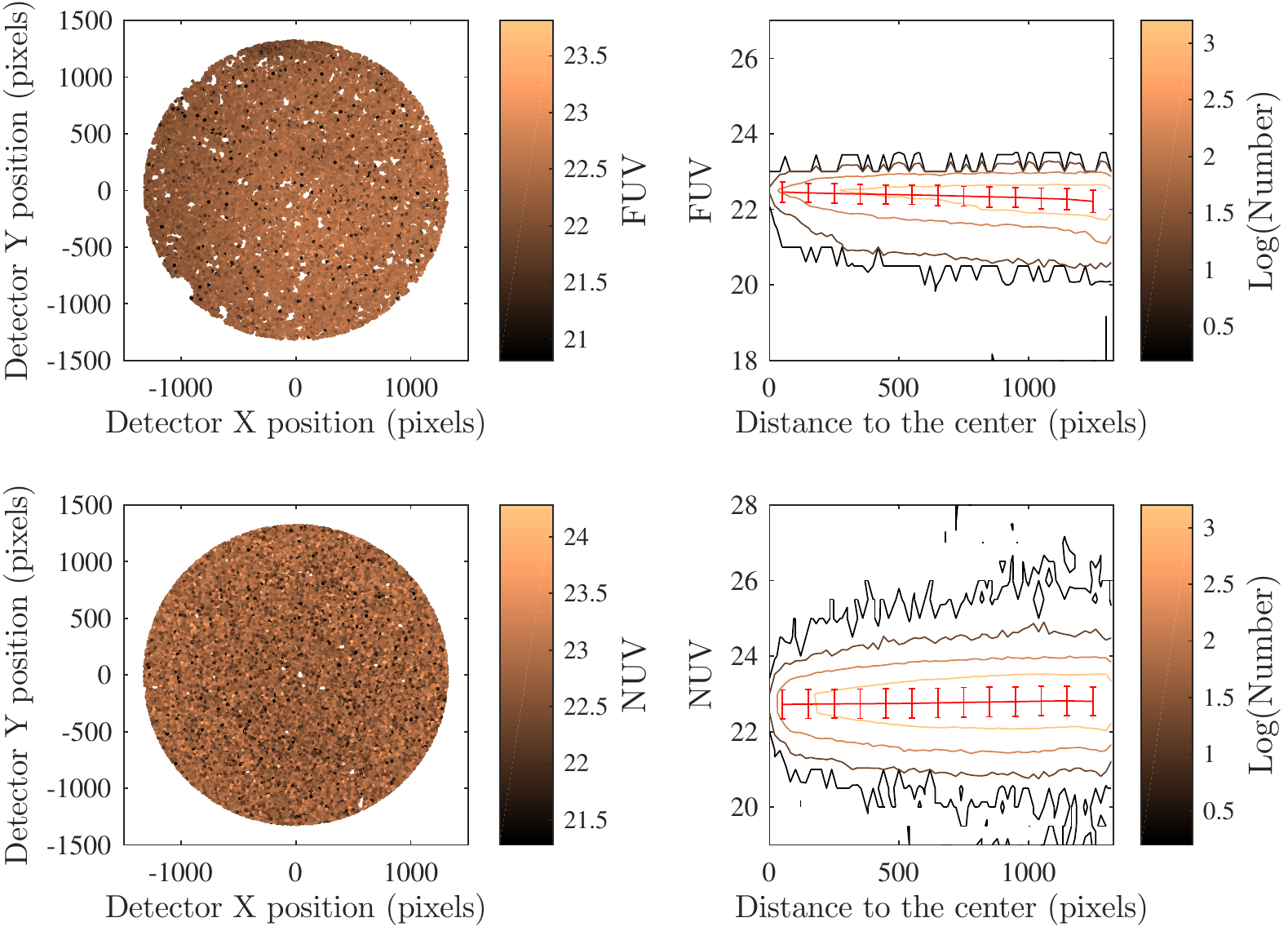}
   \caption{ The map of UV magnitudes from the detector. We accumulate all the star fields in our sample.
             Left panels: distributions of the sources with exposure time from 100 to 120 sec in the detector coordinate.
             Right panels: magnitudes as a function of the distance to the detector center. The
             mean value and the standard deviation for each bin is shown in red. The solid lines refer to the density
             contours of the magnitudes observed at a given distance from the field center.
   \label{offang}
    }
\end{figure}

We average the magnitudes of these sources in the three dimensional
bins: the Galactic coordinates $l$, $b$ and the exposure
times. The step size of each bin is 2{\arcdeg} and 1{\arcdeg} for
$l$ and $b$ respectively, and 10 seconds for exposure
times. The magnitudes in each bin of the Galactic
coordinate are then fitted with the a function similar to that in
\citet{Gezari13}:
\begin{equation}\label{e1}
m_{\rm{lim}} = -2.5~\rm{log}(5\sqrt{\it{Sky}/T_{\rm exp}})+zp,
\end{equation}
where $Sky$ is the fitted background in counts s$^{-1}$ and
$zp$ values are 18.08 and 20.08 in the FUV and NUV
respectively \citep{Morrissey07}. Nearest-neighbor
interpolation is used to estimate the magnitudes that are not valid
in some bins of the Galactic coordinates. We present the
distributions of the upper limit magnitudes on the
Galactic coordinate in Figure~\ref{gc_Up_FUVNUV}. We plot
the magnitudes with exposure times from 100 sec to 120 sec in the
map of the detector coordinate in Figure~\ref{offang}, and there is
no obvious correlation between the upper limit magnitudes
and the distances to the detector center.

The stars with low S/N are near the detection limit of $GALEX$, and
they have been used as proxies for the upper limit
magnitudes. \citet{Ansdell15} adopted an S/N of 2.
We extend the S/Ns to lower values in order to come closer to the
detection limit of $GALEX$, and use these stars to trace the
upper limit magnitudes rather than fitting the dim bound of
the UV magnitudes. We visually compare the count rate of some
randomly selected sources to that of their background, and find that
they are similar.
The magnitudes fitted from Eq.~\ref{e1} are adopted as the
upper limit magnitudes for stars in the visit and the
co-added exposures.


\subsection{Extinction}
A novel Bayesian method developed by \citet{Burnett10} and
\citet{Binney14} has been used for stars in the RAVE survey, which
has demonstrated the ability to obtain accurate distance and
extinction.
\citet{Wang16a} measured extinction and distances using this for
stars with valid stellar parameters in the first data release, and
in the second data release of the LAMOST survey (Wang. J. L.,
private communication). They used the spectroscopic
parameters \Teff, [Fe/H] and log $g$, and 2MASS photometry to
compute the posterior probability with the Bayesian method. They
adopted a three-component prior model of the Galaxy for the
distribution functions of metallicity, density, and age, in order to
construct the prior probability. They then derived the probability
distribution functions of parallaxes and extinctions. An
introduction to this technique is given in the appendixes of
\citet{Wang16a} and \citet{Wang16b}.

For stars without valid extinction measured with Bayesian method, we
use the RJCE method to estimate their extinction. This method has
shown that the extinctions are reliable for most stars. The
RJCE method derives the extinction by combining both the near and
mid infrared. We calculate A$_{K_s}$ from the equation:
\begin{equation}
\mathrm{A}_{K_s} = 0.918(H - [4.5\mu\mathrm{m}] - (H - [4.5\mu\mathrm{m}])_0),
\end{equation}
where the [4.5$\mu$m] data are the photometry of $W$2 from the $ALLWISE$ catalog \citep{Majewski11,Cutri13}.

Here ($H - [4.5\mu\mathrm{m}])_0$ is the zero point that depends on spectral types \citep{Majewski11}.
We use the BT-Cond grid \citep{Baraffe03,Barber06,Allard10} \footnote{https://phoenix.ens-lyon.fr/Grids/BT-Cond/}
of PHOENIX photospheric model \citep{Hauschildt99} in Theoretical Model Services (TMS)
\footnote{http://svo2.cab.inta-csic.es/theory/main/} to calculate
the the effective-temperature dependent zero points.
Based on the calibration provided by \citet{Gray09}, the $subclass$ in the LAMOST catalog
is used to estimate their effective temperature for stars without valid effective temperatures.
The extinction provided by \citet{Schlegel98} is adopted, if the extinction estimated from the RJCE is
larger than that of the Milky Way.

\begin{figure}
   \centering
   \includegraphics[width=0.5\textwidth]{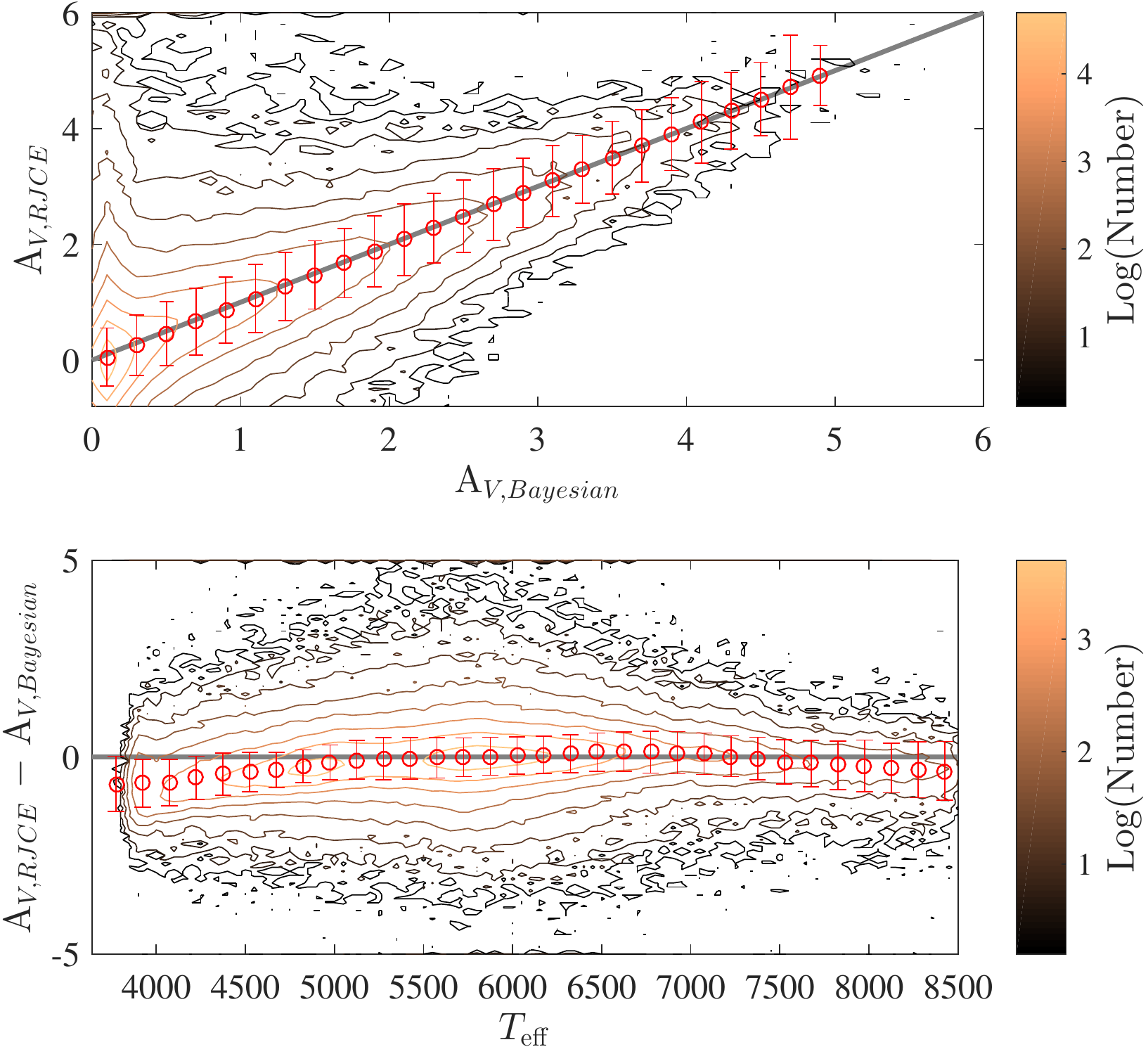}
   \caption{ The comparison between extinction derived from the Bayesian and the RJCE methods.
             Upper panel: the density contours of the extinction.
             Lower panel: the density contours of the difference as a function of effective temperature.
             In each panel an equal value is indicated with the solid grey lines.
             The solid color lines are density contours in the bin size of 0.1 magnitude
             of the extinction and 50 K of the effective temperature.
             The red circles and error bars stand for the mean values and standard deviations
             in the bins respectively.
   \label{AV1}
    }
\end{figure}

Figure~\ref{AV1} shows the comparison between the extinction derived
from the Bayesian and RJCE methods. There is no systematical
deviation between the two methods statistically, but about five
percent of the stars have $A_V$ differences larger than one.
For A and K stars, extinction from the Bayesian
method is higher than that from the RJCE within a
discrepancy of $\Delta$A$_{V}$ $\lesssim$ 0.7 mag, which is
consistent with the result of \citet{Wang16a}. Extinction
values from Bayesian method suffer uncertainty in the stellar
parameters, if they are not well constrained in the LAMOST pipeline
(\citealt{Wu11}, Sec 4.4 of \citealt{Luo15}). However, the Bayesian
methodology offers several advantages compared to RJCE as discussed
in \citet{Sale12}. The RJCE could overestimate the extinction for
stars with small $A_V$ \citep{Rodrigues14}, which can be seen in the
upper panel of Figure~\ref{AV1}.

\section{Catalog} \label{Content}

\subsection{Magnitudes}

We present the FUV and NUV counterparts from both the co-added and the visit exposures in Table~\ref{Table1}.
For stars observed but undetected by $GALEX$, we give their upper
limits in the FUV and NUV. Since the co-added exposures are combined
from some but not all the visit exposures, we flag the visit
exposures that are combined to the listed co-added exposures.

All the magnitudes in Table~\ref{Table1} do not reach the limits of saturation
\footnote{http://www.galex.caltech.edu/researcher/} before the extinction correction.
The saturation effect is insignificant in our sample, since about 0.3\% and 0.1\% of the stars
in the FUV and NUV reach the more conservative limits of $m_{\rm FUV}$ $<$ 12.6 and $m_{\rm NUV}$ $<$ 13.9
\citep{Morrissey07,Findeisen11}.

We check the distributions of exposure time in our sample. The bimodal distributions are dominated
by images from the all-sky imaging survey (AIS) and the medium imaging survey (MIS).

\begin{figure}
   \centering
   \includegraphics[width=0.5\textwidth]{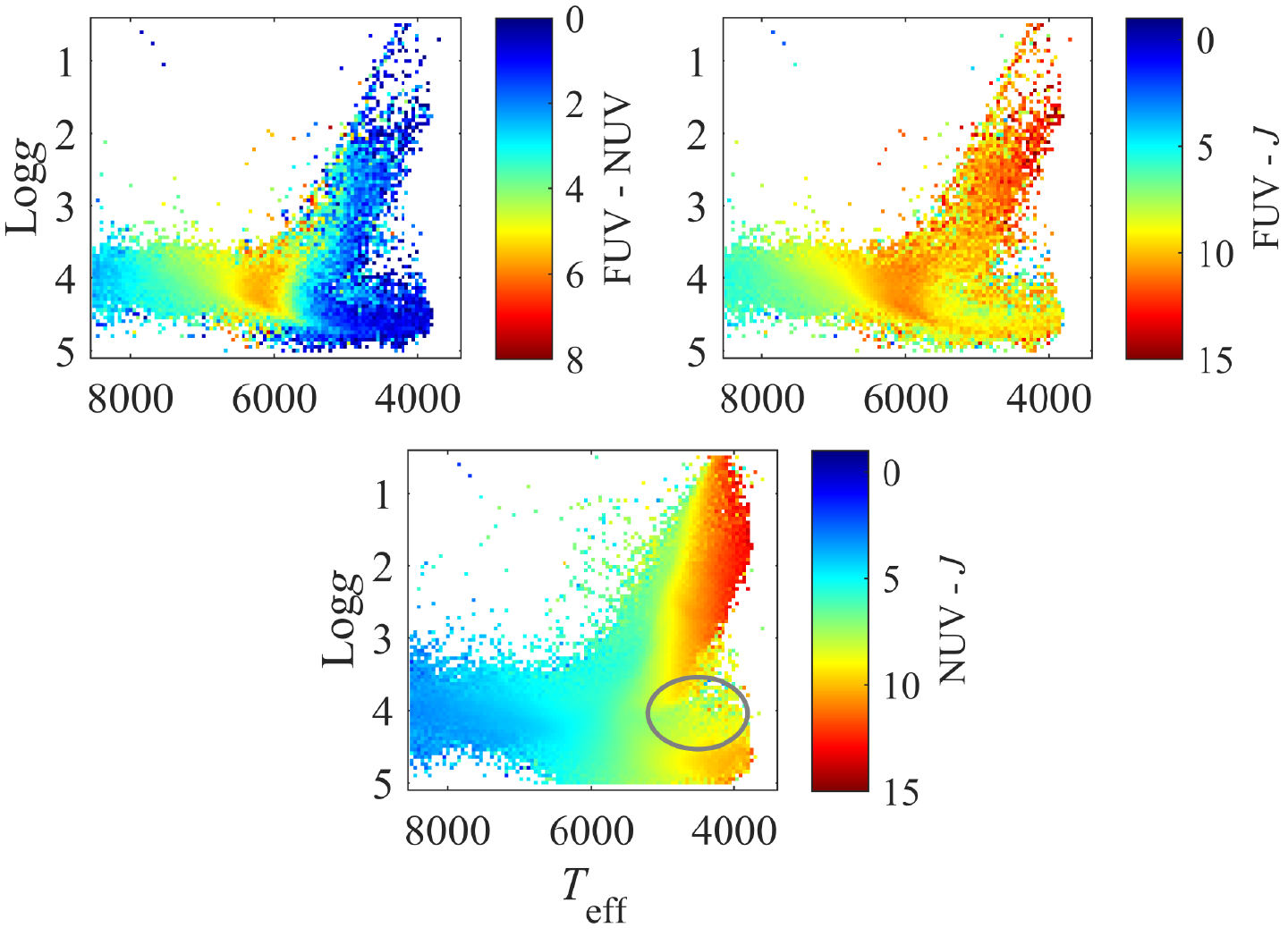}
   \caption{The distributions of stars in the Hertzsprung-Russell-like (HR-like) diagram.
            The color bars represent the colors of the (FUV $-$ NUV), the (FUV $-$ $J$),
            and the (NUV $-$ $J$) in the bins of log $g$ and {\Teff}.
            The K stars with bluer (UV $-$ IR) colors than other K stars are circled in grey.
   \label{HR_Color_m}}
\end{figure}

We calculate the mean magnitudes from the average fluxes of the visit exposures, and present the colors'
distributions in Figure~\ref{HR_Color_m}. The (FUV $-$ NUV) is not only
very sensitive to the young massive stars \citep{Kang09}, but also a useful indicator of stellar
activities for low-mass stars \citep{Welsh06,France16}.
Both K giants and dwarfs have blue (FUV $-$ NUV),
probably due to the strong FUV line emission from upper atmosphere layers primarily in emission lines
of \Ion{C}{II}, \Ion{Si}{IV}, and \Ion{C}{IV} \citep{Jones16}.
The stars of {\Teff} $\sim$ 6000 K are redder than others, forming a red valley in the HR-like diagram.
On one hand, their colors are redder than earlier-type stars due to the lower effective temperatures.
On the other hand, their FUV emission from upper regions of their atmosphere is weaker than
that of later-type stars probably due to their weaker upper atmosphere emission.

The infrared bands, which are dominated by flux from the
photosphere, are insensible to upper atmosphere emission.
The very blue (FUV $-$ $J$) colors for K dwarfs are due to
emission from upper atmosphere layers. The (NUV $-$ $J$) colors for
the K dwarfs are not very blue because the NUV emission is mostly
photospheric. The colors of (FUV $-$ $J$) and (NUV $-$ $J$) decline
with increasing {\Teff}, and dwarfs are bluer than giants.
We find that the K stars with {\Teff} $\sim$ 4500 K, log $g$ $\sim$
4 (the grey circles in Figure \ref{HR_Color_m}), are bluer than
other K stars.
The bluer (UV $-$ IR) colors and smaller gravities might be due to their rapid
rotations \citep{Stelzer13} or the interactions with spectrally unresolved companions \citep{Ansdell15}.


\begin{turnpage}
\begin{deluxetable*}{cccccccrlllllllllll}
\tablecaption{UV Data in Our Catalog \label{Table1}}
\tablehead{\colhead{LID} & \colhead{Type}  & \colhead{\Teff} & \colhead{log $g$}           & \colhead{$A_{\rm FUV}$} & \colhead{$A_{\rm NUV}$}& \colhead{f$_{E}$} & \colhead{MJD}   & \colhead{FUV$_{\rm{l, v}}$}   & \colhead{NUV$_{\rm{l, v}}$} & \colhead{f$_{C}$} & \colhead{FUV$_{\rm v}$} & \colhead{NUV$_{\rm v}$} & \colhead{f$_{\rm{F}}$} & \colhead{f$_{\rm{N}}$} & \colhead{FUV$_{\rm l}$} & \colhead{NUV$_{\rm l}$}& \colhead{FUV}   & \colhead{NUV}  \\
           \colhead{}      & \colhead{}      & \colhead{(K)} & \colhead{(cm s$^{-2}$)}     & \colhead{(mag)} & \colhead{(mag)}                & \colhead{}        & \colhead{(day)} & \colhead{(mag)}                   & \colhead{(mag)}                 & \colhead{}        & \colhead{(mag)}           & \colhead{(mag)}           & \colhead{}             & \colhead{}             & \colhead{(mag)}           & \colhead{(mag)}          & \colhead{(mag)} & \colhead{(mag)}\\
           \colhead{(1)}   & \colhead{(2)}   & \colhead{(3)} & \colhead{(4)}   & \colhead{(5)}    & \colhead{(6)}     & \colhead{(7)}   & \colhead{(8)}                     & \colhead{(9)}                   & \colhead{(10)}    & \colhead{(11)}            & \colhead{(12)}            & \colhead{(13)}         & \colhead{(14)}         & \colhead{(15)}            & \colhead{(16)}           & \colhead{(17)}  & \colhead{(18)} & \colhead{(19)}
           }
\startdata
    607005 &    F5 &  6603 & 4.22 &  0.37 &  0.55 & 1 & 3236.0179 & 21.26 & 21.90 &  1 &                   & 17.69 $\pm$  0.06        &   & 0 & 21.70 & 22.25 &                   & 17.61 $\pm$  0.05        \\
           &       &       &      &       &       &   & 3241.4768 & 21.34 & 21.90 &  1 &                   & 17.61 $\pm$  0.06        &   & 0 & 22.98 & 23.54 & 22.28 $\pm$  0.30 & 17.58 $\pm$  0.01        \\
           &       &       &      &       &       &   & 4664.2107 & 22.98 & 23.54 &  2 & 22.28 $\pm$  0.30 & 17.58 $\pm$  0.01        & 0 & 0 &       &       &                   &                          \\
    607006 &    M0 &  3759 &      &  0.39 &  0.58 & 0 & 3236.0179 & 21.24 & 21.87 &  0 &                   &                          &   &   & 21.67 & 22.22 &                   &                          \\
           &       &       &      &       &       &   & 3241.4768 & 21.32 & 21.87 &  0 &                   &                          &   &   & 22.96 & 23.51 &                   &                          \\
           &       &       &      &       &       &   & 4664.2107 & 22.96 & 23.51 &  0 &                   &                          &   &   &       &       &                   &                          \\
    607007 &    K0 &  5330 & 4.56 &  0.36 &  0.53 & 1 & 3236.0179 & 21.27 & 21.92 &  0 &                   &                          &   &   & 21.71 & 22.28 &                   &                          \\
           &       &       &      &       &       &   & 3241.4768 & 21.35 & 21.92 &  0 &                   &                          &   &   & 23.00 & 23.56 &                   &                          \\
           &       &       &      &       &       &   & 4664.2107 & 23.00 & 23.56 &  0 &                   &                          &   &   &       &       &                   &                          \\
    607009 &    G5 &  4866 & 2.50 &  0.40 &  0.59 & 1 & 3990.2005 & 21.80 & 22.22 &  1 &                   & 21.15 $\pm$  0.32        &   & 0 & 21.80 & 22.22 &                   & 21.15 $\pm$  0.32        \\
    607011 &    F9 &  6010 &      &  0.00 &  0.00 & 0 & 3236.0179 & 21.63 & 22.45 &  0 &                   & 20.37 $\pm$  0.31        &   & 1 & 22.07 & 22.81 &                   &                          \\
           &       &       &      &       &       &   & 3241.4768 & 21.71 & 22.45 &  0 &                   & 20.77 $\pm$  0.27        &   & 1 & 22.07 & 22.81 &                   &                          \\
           &       &       &      &       &       &   & 3990.2027 & 22.07 & 22.81 &  0 &                   &                          &   &   &       &       &                   &                          \\
    607013 &    G3 &  5955 & 4.34 &  0.27 &  0.40 & 1 & 3236.0179 & 21.50 & 22.06 &  0 &                   & 19.06 $\pm$  0.13        &   & 0 & 21.94 & 22.41 &                   &                          \\
           &       &       &      &       &       &   & 3241.4768 & 21.58 & 22.06 &  0 &                   & 19.50 $\pm$  0.22        &   & 1 & 23.22 & 23.70 &                   & 19.18 $\pm$  0.03 $^{*}$ \\
           &       &       &      &       &       &   & 4664.2107 & 23.22 & 23.70 &  1 &                   & 19.18 $\pm$  0.03 $^{*}$ &   & 0 &       &       &                   &                          \\
    607015 &    G3 &  5881 & 4.50 &  0.35 &  0.51 & 1 & 3236.0179 & 21.42 & 21.94 &  0 &                   & 20.74 $\pm$  0.44        &   & 1 & 21.86 & 22.30 &                   &                          \\
           &       &       &      &       &       &   & 3241.4768 & 21.50 & 21.94 &  0 &                   & 20.54 $\pm$  0.33        &   & 0 &       &       &                   &                          \\
    607016 &    F5 &  6374 & 4.21 &  0.35 &  0.51 & 1 & 3990.2005 & 21.86 & 22.30 &  1 &                   & 18.50 $\pm$  0.07        &   & 0 & 21.86 & 22.30 &                   & 18.50 $\pm$  0.07        \\
    607018 &    K5 &  4400 &      &  0.00 &  0.00 & 0 & 3990.2005 & 22.20 & 22.81 &  0 &                   &                          &   &   & 22.20 & 22.81 &                   &                          \\
    607019 &    K3 &  4718 & 3.65 &  0.39 &  0.57 & 1 & 3990.2005 & 21.82 & 22.24 &  0 &                   &                          &   &   & 21.82 & 22.24 &                   &                          \\
    607021 &    K7 &  4104 & 4.75 &  0.18 &  0.26 & 1 & 3236.0179 & 21.46 & 22.19 &  0 &                   &                          &   &   & 21.89 & 22.54 &                   &                          \\
           &       &       &      &       &       &   & 3241.4768 & 21.53 & 22.19 &  0 &                   &                          &   &   & 23.18 & 23.83 &                   &                          \\
           &       &       &      &       &       &   & 4664.2107 & 23.18 & 23.83 &  0 &                   &                          &   &   &       &       &                   &
\enddata

\tablecomments{All magnitudes are corrected for extinction.
The FUV and NUV magnitudes are extracted from the catalogs
of $GALEX$ GR6/GR7 and corrected for extinction.  Column (1): IDs
in the LAMOST catalog. Column (2): Spectral types. Column (3):
Effective temperatures in the AFGK catalog or estimated with the
stellar subclasses. Column (4): Surface Gravity. Column (5)-(6):
Extinction in the UV. Column (7): Flags of extinction. f$_E$ = 0,
extinction calculated with the RJCE method; f$_E$ = 1, extinction
calculated with Bayesian method. Column (8): JD $-$ 2450000. Column
(9)-(10): Upper limit magnitudes from visit exposures.
Column (11): Flags mark the visit exposures combined into
the corresponding co-added exposures for the detected stars. Column
(12)-(13): Magnitudes for the stars detected from visit exposures.
We mark the stars with asterisks when there are more than one
counterpart within 5\arcsec in the same visit exposure. Column
(14)-(15): Flags for the stars with flag\_artifact $>$ 1 in the FUV
and NUV, when f = 1. Column (16)-(17): Upper limit
magnitudes from co-added exposures. Column (18)-(19): Magnitudes for
the stars detected from co-added exposures. We mark the stars with
asterisks when there are more than one counterpart within 5\arcsec
in the same co-added exposure. }
\end{deluxetable*}
\end{turnpage}


\subsection{UV Excesses}\label{NorR}

The UV excess (Eq.~\ref{EUV}) and the normalized UV excess (Eq.~\ref{RUV}) describe the
distributions of UV flux with respect to a basal value:
\begin{equation}\label{EUV}
E_\mathrm{UV} = m_\mathrm{UV,obs}-m_\mathrm{UV,ph},
\end{equation}
\begin{equation}\label{RUV}
R'_\mathrm{UV} = \frac{f_\mathrm{UV,obs}-f_\mathrm{UV,ph}}{f_\mathrm{bol}},
\end{equation}
where $f_\mathrm{UV,obs}$ ($m_\mathrm{UV,obs}$) and $f_\mathrm{UV,ph}$ ($m_\mathrm{UV,ph}$)
are the observed UV flux (magnitude)
and the photospheric flux (magnitude) in the same UV band. The $f_\mathrm{bol}$ is
the bolometric flux calculated from the effective temperature
\citep{Findeisen11,Stelzer13}.

We estimate the photospheric flux in the FUV and NUV with the
BT-Cond grid. All synthetic magnitudes in the FUV, NUV, and $J$
bands are extracted with the help of TMS. We construct a theoretical
grid for interpolation, and the dimensions are effective
temperatures, surface gravities and metallicities. The scaling
factor $(\frac{R_*}{d})^2$, the squared ratio between the stellar
radius and the distance, is calculated from the (UV $-$ $J$) color
predicted from the synthetic model. The scaling factor is
used to convert the observed UV flux to that from the stellar
surface \citep{Stelzer13}. We only compute the $E_\mathrm{UV}$ and
$R'_\mathrm{UV}$ for stars with valid stellar parameters, and
estimate their corresponding uncertainties from the errors
in magnitudes, \Teff, log $g$, and [Fe/H].

About 11\% and 22\% of the $R'_\mathrm{UV}$ are below zero in the
FUV and NUV, respectively. In these cases, the normalized UV
excesses are insignificant and probably dominated by uncertainties
in the extinction and the model interpolation. The uncertainties of
the extinction are from the errors in the stellar parameters, when
the extinction is given with the Bayesian method.  We present stars
with $R'_\mathrm{UV}$ $>$ 0 in Table~\ref{Table2}.

\begin{figure}
   \centering
   \includegraphics[width=0.45\textwidth]{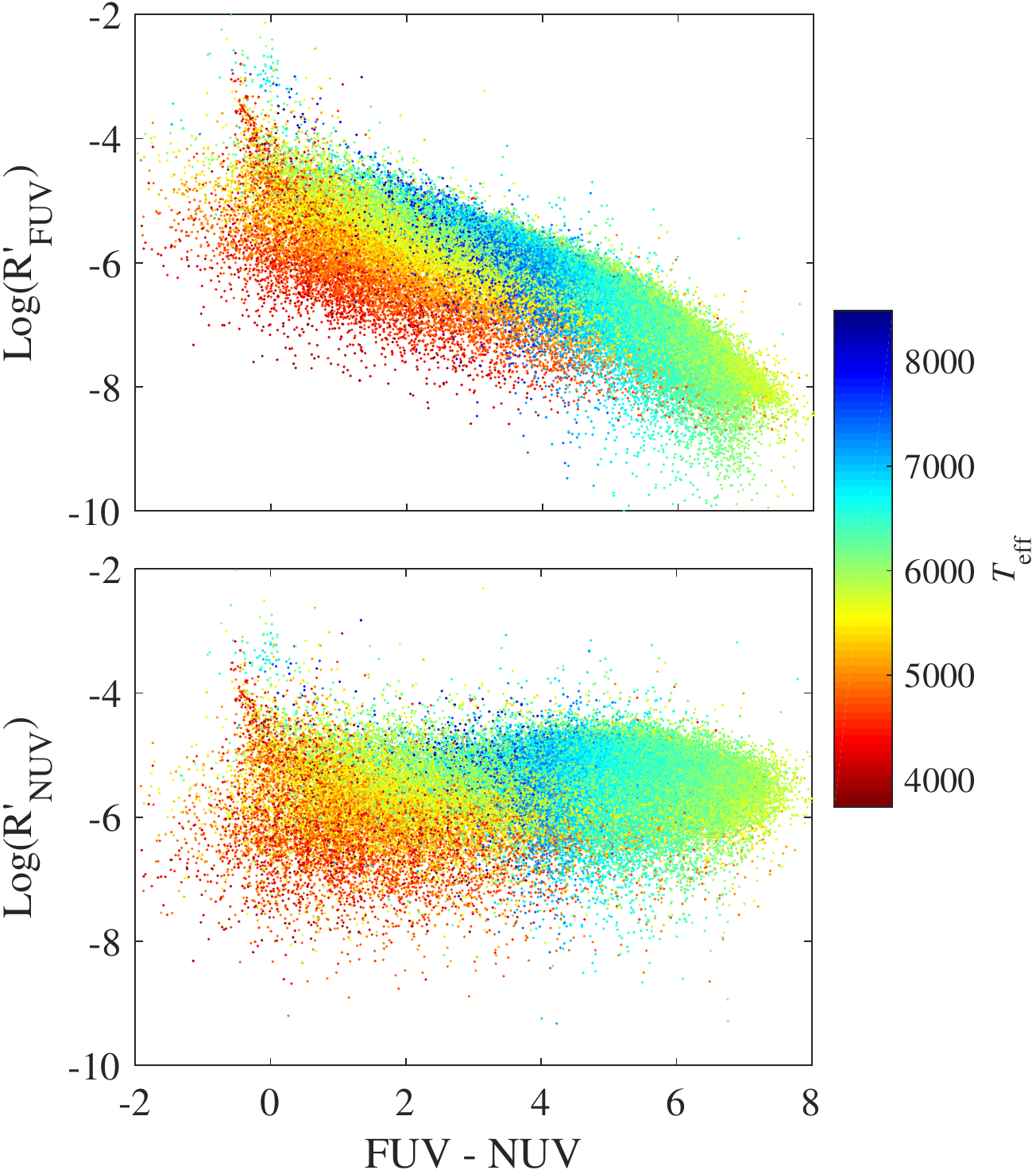}
   \caption{ The log(\Rfuv) (the upper panel) and the log(\Rnuv) (the lower panel) as functions of the (FUV $-$ NUV). The colored points represent the effective temperature
   as shown by the color bar to the right. A clear linear relation can be seen in the distribution of log(\Rfuv) vs. the (FUV $-$ NUV), but there is no such relation
   in the NUV.
   \label{fuvnuv_R_m}
    }
\end{figure}

The color excess (FUV $-$ NUV) suffers from less uncertainties than
the UV excesses, and could shed light on the stellar activity
\citep{Shkolnik13,Smith14}. The normalized FUV excess as a function
of (FUV $-$ NUV) is shown in Figure~\ref{fuvnuv_R_m}, in which the
anti-correlation can be fitted with a linear relation, \footnotesize
\begin{equation}
\mathrm{Log}(\Rfuv) = (-0.380 \pm 0.001) \times (\mathrm{FUV} - \mathrm{NUV}) - (4.85 \pm 0.01).
\end{equation}
\normalsize
It is suggested that the (FUV $-$ NUV) could trace the {\Rfuv}.
The stars with stronger upper atmosphere emission have larger \Rfuv.
These stars are mostly K stars with (FUV $-$ NUV) near zero.
The correlation implies that such emission is more obvious in the FUV than in the NUV.
Therefore, the {\Rnuv} shows no obvious correlation with the (FUV $-$ NUV) in Figure~\ref{fuvnuv_R_m}.
There is an effective temperature dependence in the {\Rfuv} plot. The dependence may be due
to the different FUV origination. The FUV of the early-type stars is mainly from the photospheres, while that of the
late-type stars is mainly from the upper atmospheres.

\begin{figure}
   \centering
   \includegraphics[width=0.5\textwidth]{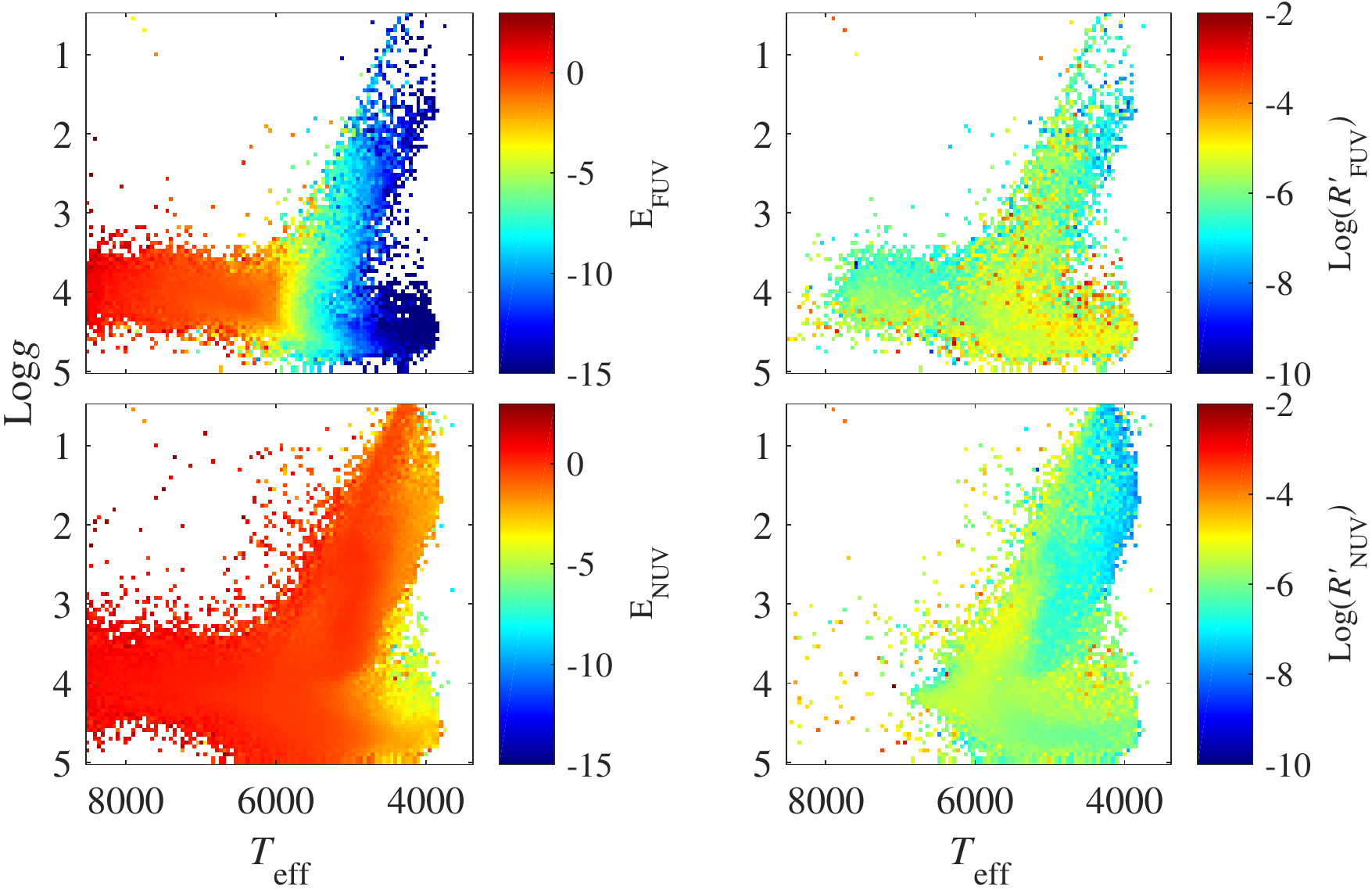}
   \caption{The UV and the normalized UV excesses in the HR-like diagram.
            Left panels: the distributions of FUV and NUV excesses in the bins of
                          {\Teff} and log $g$ with a bin size of 50 K and 0.05 dex.
            Right panels: the distributions of the normalized FUV and NUV excesses.
                          For stars with high {\Teff}, the normalized UV excesses are invalid due to
                          the negative mean values in some bins.
   \label{HR_E_R_m}
    }
\end{figure}

We present the color-coded distributions of the UV and the
normalized UV excesses in Figure~\ref{HR_E_R_m}, in which the
excesses depend on {\Teff} and log $g$. The {\Efuv} increases with
increasing {\Teff}. The {\Enuv} has a similar distribution
but with a smaller increasing level. This indicates that
low mass stars have strong UV excesses, especially in the FUV.

The {\Rfuv} is higher than the {\Rnuv} for stars with {\Teff}
$\lesssim$ 6000 K, since their stellar activities might be more
efficient at powering the the line emission in the
FUV than the continuum in the NUV \citep{Jones16}. The
$R'_\mathrm{UV}$ of giants is lower than that of dwarfs, because the
upper atmosphere emission of giants are buried beneath
their atmospheres \citep{Holzwarth01,Nucci14}. There are some bins
with average $R'_\mathrm{UV}$ below zero in Figure~\ref{HR_E_R_m},
which results in the empty areas of early-type stars.

\begin{deluxetable*}{rrrrrrrrrrrrrrr}
\tablecaption{UV Parameters in Our Catalog \label{Table2}}
\tablehead{\colhead{LID} & \colhead{\Efuv} & \colhead{$eE_{\mathrm{FUV}}$ } & \colhead{{\Enuv}} & \colhead{$eE_{\mathrm{NUV}}$ }  & \colhead{log {\Rfuv}} & \colhead{log $eR'_{\mathrm{FUV}}$ } & \colhead{log {\Rnuv}} & \colhead{log $eR'_{\mathrm{NUV}}$ } & \colhead{$\sigma_{\mathrm{int,FUV}}$} & \colhead{$\sigma_{\mathrm{int,NUV}}$} & \colhead{$S_{\mathrm{d,FUV}}$} & \colhead{$S_{\mathrm{d,NUV}}$} \\
           \colhead{(1)}   & \colhead{(2)}   & \colhead{(3)}           & \colhead{(4)}   & \colhead{(5)}                  & \colhead{(6)}   & \colhead{(7)}           & \colhead{(8)}   & \colhead{(9)}                  & \colhead{(10)}     & \colhead{(11)} & \colhead{(12)}     & \colhead{(13)}
           }
\startdata
    101077& $-$2.65& 1.51&    1.25& 0.61& $-$5.37& $-$5.72&        &        &    &    &    &0.34  \\
    101150& $-$8.80& 0.91& $-$0.54& 0.49& $-$5.08& $-$5.46& $-$5.74& $-$5.86&    &0.52&    &1.54  \\
    103032& $-$0.29& 0.76& $-$0.44& 0.23& $-$8.07& $-$7.89& $-$5.51& $-$5.86&0.41&0.03&    &      \\
    104056& $-$0.47& 0.35& $-$0.35& 0.16& $-$7.19& $-$7.39& $-$5.20& $-$5.60&0.69&0.02&0.11&0.01  \\
    106049& $-$5.74& 1.16& $-$0.11& 0.29& $-$5.49& $-$6.37& $-$6.11& $-$5.47&0.33&0.04&0.93&0.12
\enddata
\tablecomments{
Column (1): Object IDs in the LAMOST catalog.
Column (2): Excesses in the FUV.
Column (3): Errors of {\Efuv}.
Column (4): Excesses in the NUV.
Column (5): Errors of {\Enuv}.
Column (6): Normalized FUV Excesses.
Column (7): Errors of {\Rfuv}.
Column (8): Normalized NUV Excesses.
Column (9): Errors of {\Rnuv}.
Column (10): Intrinsic variabilities in the FUV.
Column (11): Intrinsic variabilities in the NUV.
Column (12): Structure functions on timescales of days in the FUV.
Column (13): Structure functions on timescales of days in the NUV.
}
\end{deluxetable*}

\subsection{Variability Statistics}\label{VS}

\begin{figure}
   \centering
   \includegraphics[width=0.45\textwidth]{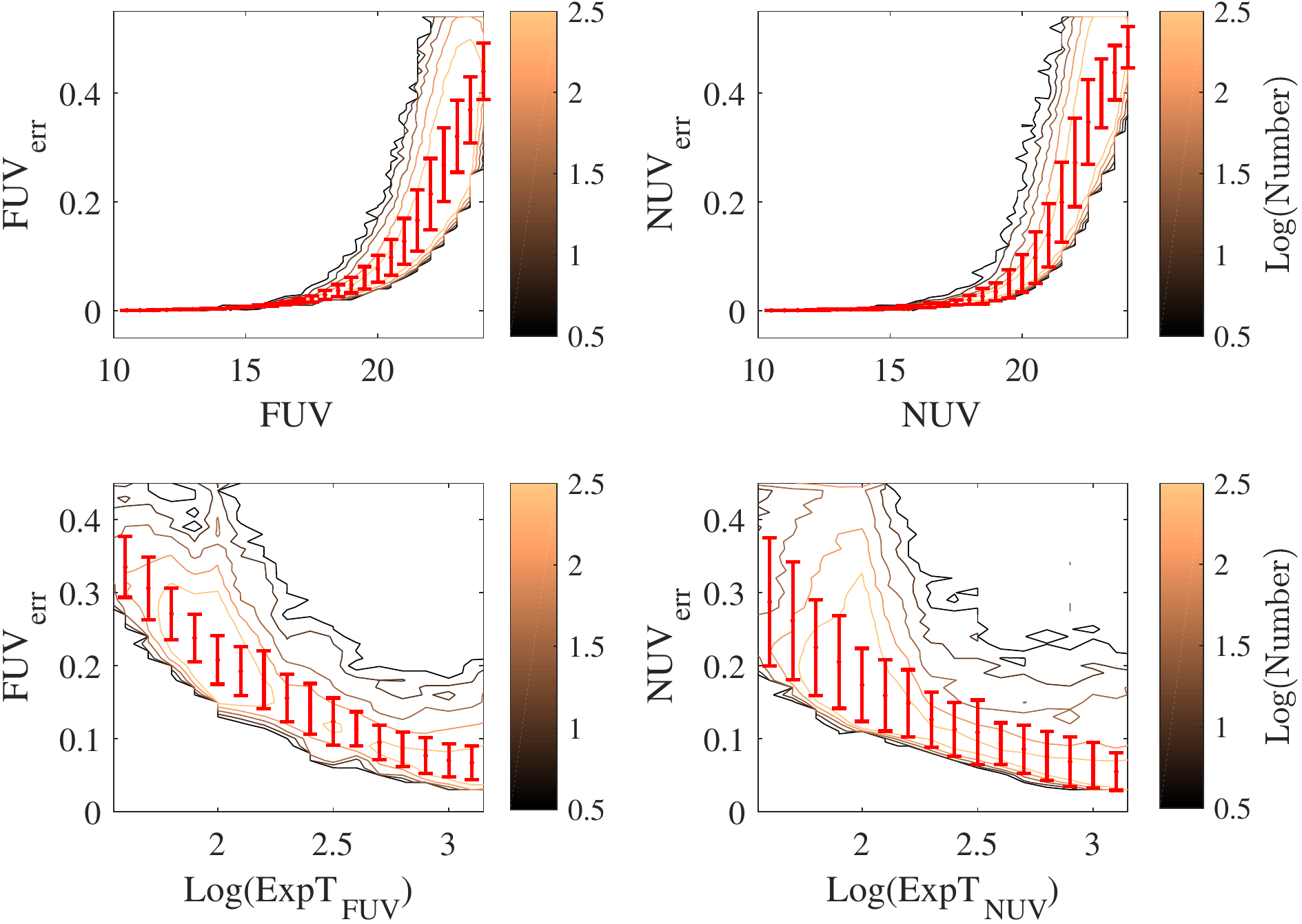}
   \caption{The uncertainties as functions of the magnitudes and the exposure times in the FUV (left panels) and the
            NUV (right panels).  The solid color lines are
            density contours with a bin size of 0.5 magnitudes, 0.01 uncertainties, and 0.1 seconds in the log scale.
            The red points and error bars stand for the mean values and standard deviations in the bins respectively.
   \label{err}}
\end{figure}

The rms scatter has been used to describe the intrinsic variability in optical and UV bands
\citep{Sesar07,Gezari13},
\begin{equation}
\sigma_\mathrm{int} = \sqrt{\Sigma^2-\langle\xi\rangle^2},
~\mathrm{where}~\Sigma = \sqrt{\frac{1}{n-1}\sum^n_{i=1}(m_i-\langle{m}\rangle)^2}.
\end{equation}
In order to estimate the mean photometric error $\langle\xi\rangle$,
we extract over six million objects in the FUV and NUV bands. The
mean photometric error depends on both the magnitude and the
exposure time in Figure~\ref{err}. We construct a two
dimensional grid with bin size of ${\Delta}m$ = 0.5 magnitude and
$\Delta$log(ExpT) = 0.1 second.

\begin{figure}
   \centering
   \includegraphics[width=0.5\textwidth]{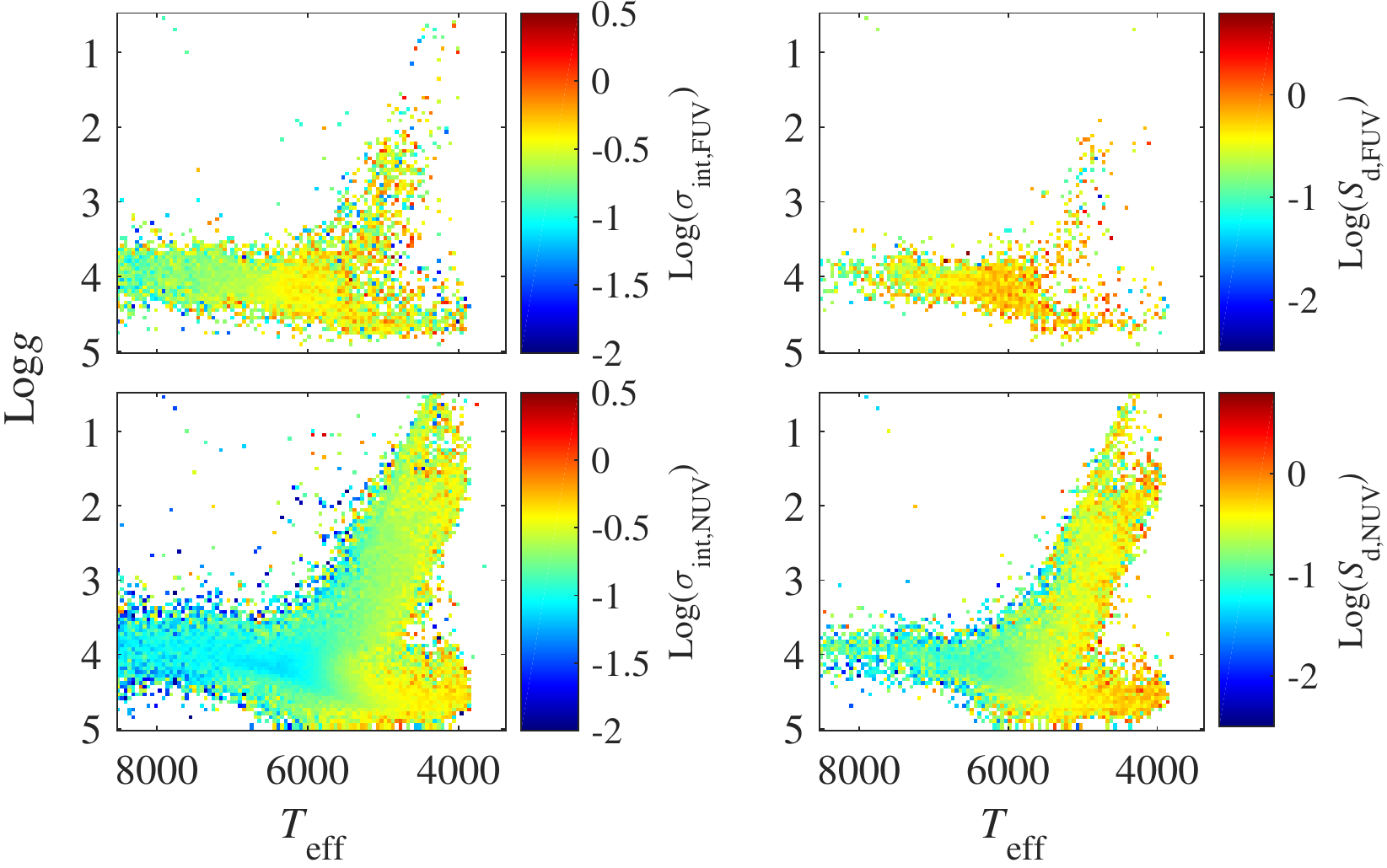}
   \caption{ The parameters of variations in the HR-like diagram.
   Left panels: the distributions of the intrinsic variations in the bins of log $g$ and {\Teff}.
   Right panels: same as the left panels but for the structure functions, $S_{\rm d}$ in the FUV and NUV.
   \label{HR_S_S}}
\end{figure}

We then obtain mean photometric error with the interpolation, and
calculate the rms scatter $\sigma_{\rm{int}}$ of the stars detected
in at least two visit exposures. It yields 46,022 and
700,039 stars with available $\sigma_{\rm{int}}$ in the FUV and NUV,
respectively. The distributions of $\sigma_ {\rm{int}}$ are
presented in the left panels of Figure~\ref{HR_S_S}. The intrinsic
variability declines with the increase of {\Teff}, implying
that late-type stars tend to have strong variabilities. The
$\sigma_{\rm int}$ values in the FUV are generally larger
than those in the NUV, probably due to the higher
amplitudes of light curves in the shorter wavelength
\citep{Wheatley12}.

Another parameter to characterize the variability is the structure function (see \citealt{di96} for details),
\begin{equation}
V(\Delta{t}) = \sqrt{\frac{\pi}{2}\langle|\Delta{m_{ij}}|\rangle^{2}_{\Delta{t}}-\langle\sigma^{2}_i+\sigma^{2}_j\rangle  },
\end{equation}
where brackets stand for averages for all pairs of points on the
light curve of an individual source with $i$ $<$ $j$,
$\Delta{m_{ij}}=m(t+\Delta{t})-m(t)$ and $t_j-t_i=\Delta{t}$. We
calculate the structure functions on the time scale of days,
$S_{\rm{d}}$, from the characteristic-time bins of
$V(\Delta{t})$. The $S_{\rm{d}}$ is defined as the maximum
value of the structure function evaluated for
$\Delta{t_{2d}}=2\pm0.5$ d, $\Delta{t_{4d}}=4\pm0.5$ d,
$\Delta{t_{6d}}=6\pm0.5$ d, $\Delta{t_{8d}}=8\pm0.5$ d
\citep{Gezari13}. This value is also presented in
Table~\ref{Table2}. The distributions of $S_{\rm{d}}$ are shown in
the right panels of Figure~\ref{HR_S_S}.
The distributions of $\sigma_{\rm{int}}$ and $S_{\rm{d}}$ are
similar. The stars with lower {\Teff} have higher
variabilities, which might be naturally explained by the strong
upper atmosphere emission of late-type stars.

\subsection{False Positives}

The stars can appear UV-luminous for some reasons not related to
their upper atmosphere emission or photospheric
radiation. These cases are so called false positives, which
are false matched background active galactic nuclei (AGNs) within
$\sim$5\arcsec. Another kind of possible false positive is
binaries composed of non-degenerate stars and white dwarfs.
These binaries are unresolved by spectra from the
LAMOST facility. The wrongly selected UV counterpart should
also be considered as false positives, when there are more than one
counterpart within {5\arcsec} in the same exposure.

We estimate the probability of false positives from background AGNs
by matching false-star positions within a random arcminute of their
true positions with the $GALEX$ archive data \citep{Jones16}. It
yields about a 0.14\% chance of matching for given a random
position, and about 0.25\% for our sample given over 55\% of the
LAMOST stars are detected in $GALEX$.

\begin{figure}
   \centering
   \includegraphics[width=0.45\textwidth]{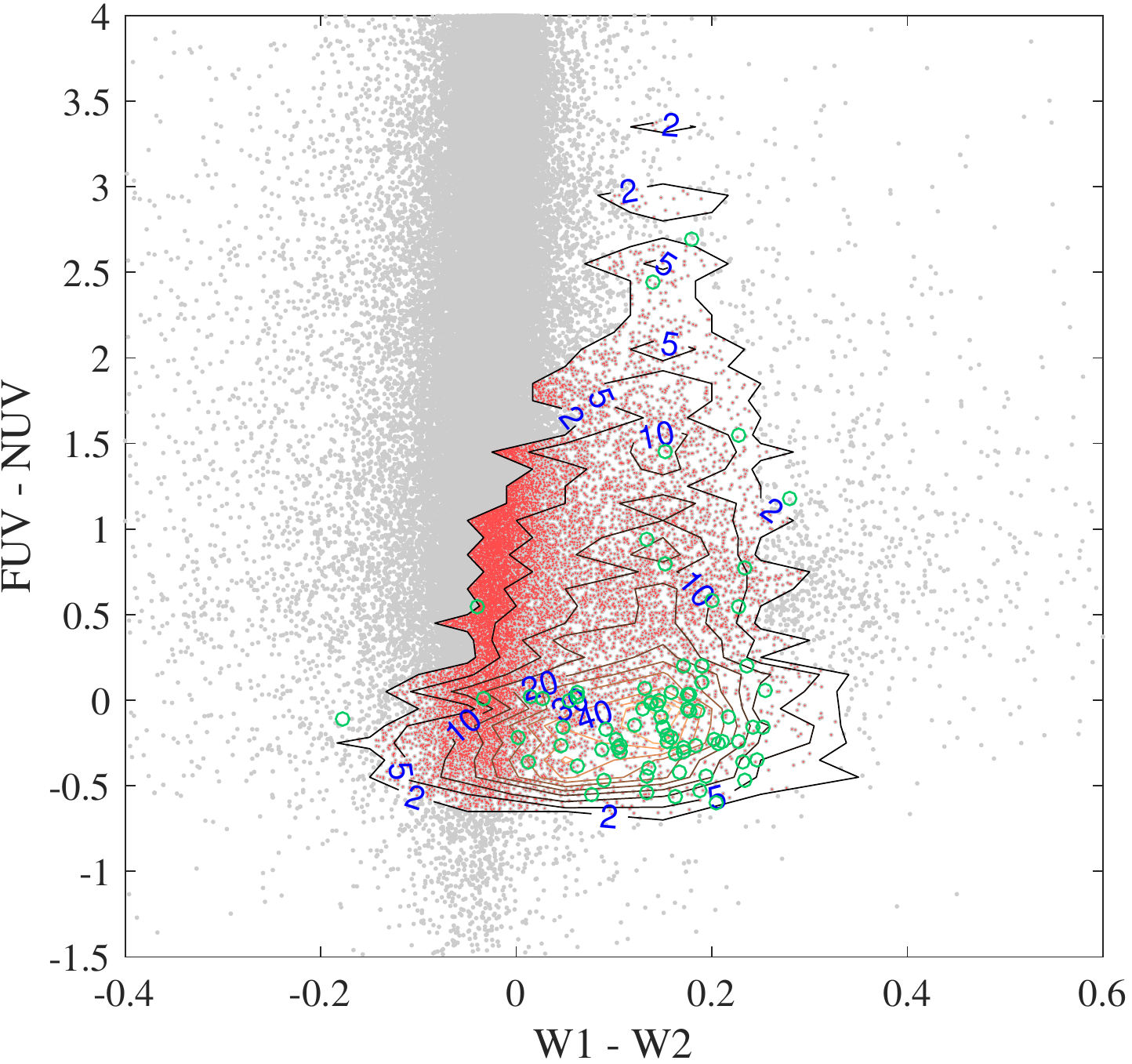}
   \caption{The diagram of (FUV $-$ NUV) vs. (W1 $-$ W2) for the stars in our sample and the WDMS binaries.
   The solid lines are density contours in the
   bin size of 0.1 magnitude for WDMS binaries of \citet{Rebassa13}. The densities are marked as blue
   numbers along the contours.
   The WDMS binaries of \citet{Ren14} are shown as green circles.
   The stars in our sample located inside the density contour
   of two per bin are shown as red points, and others are shown in grey.
   \label{WDMD2_m}}
\end{figure}

The catalog of \citet{Rebassa13} is used to select the
potential white dwarf main-sequence (WDMS) binaries in our sample.
We follow the same procedure described in Section~\ref{data} to find
the UV counterparts of these WDMS binaries. We find that the
combination of UV and IR could separate WDMS binaries, since they
cluster at FUV $-$ NUV $\sim$ 0 and W1 $-$ W2 $\sim$ 0.14 in
Figure~\ref{WDMD2_m}. There are 17,428 stars in our sample, $\sim$
0.56\%, located inside the density contour of two per bin, which are
potential WDMS-binary candidates unresolved by LAMOST
(Table~\ref{Table3}).

There are 13,845 stars having more than one counterpart within
{5\arcsec} in the same co-added or visit exposures. If they are all
wrongly selected as UV counterparts, the occurrence of false
positives is 0.45\%. The overall occurrence of these false positives
is below 1.3\% in our sample.

\begin{deluxetable}{ccrcrcc}
\tablecaption{The WDMS Candidates in Our Catalog \label{Table3}}
\tablehead{\colhead{LID} & \colhead{FUV} & \colhead{NUV} & \colhead{W1} & \colhead{W2}  \\
           \colhead{(1)}   & \colhead{(2)}   & \colhead{(3)}           & \colhead{(4)}   & \colhead{(5)}
           }
\startdata
    102042 & 21.56 $\pm$ 0.13&20.64 $\pm$ 0.08 & 13.02 $\pm$ 0.02 & 12.92 $\pm$ 0.03\\
    102173 & 21.54 $\pm$ 0.44&21.86 $\pm$ 0.19 & 14.57 $\pm$ 0.03 & 14.62 $\pm$ 0.06\\
    103165 & 23.26 $\pm$ 0.30&23.82 $\pm$ 0.36 & 14.25 $\pm$ 0.03 & 14.13 $\pm$ 0.04\\
    104029 & 21.66 $\pm$ 0.31&21.74 $\pm$ 0.18 & 13.32 $\pm$ 0.03 & 13.36 $\pm$ 0.03\\
    104250 & 21.34 $\pm$ 0.06&21.06 $\pm$ 0.02 & 12.64 $\pm$ 0.02 & 12.60 $\pm$ 0.03\\
    106053 & 23.04 $\pm$ 0.44&21.31 $\pm$ 0.01 & 14.65 $\pm$ 0.03 & 14.61 $\pm$ 0.07\\
    106057 & 22.13 $\pm$ 0.43&22.56 $\pm$ 0.31 & 13.27 $\pm$ 0.03 & 13.14 $\pm$ 0.03\\
    106063 & 22.32 $\pm$ 0.19&22.67 $\pm$ 0.07 & 14.45 $\pm$ 0.03 & 14.50 $\pm$ 0.06\\
    106241 & 22.32 $\pm$ 0.19&20.85 $\pm$ 0.02 & 15.89 $\pm$ 0.06 & 15.73 $\pm$ 0.16\\
    107224 & 21.06 $\pm$ 0.46&20.13 $\pm$ 0.07 & 13.24 $\pm$ 0.02 & 13.25 $\pm$ 0.03

\enddata
\tablecomments{All magnitudes are corrected for extinction. Column
(1): Object IDs in the LAMOST catalog. Column (2): Mean magnitudes
calculated from all the magnitudes in the visit exposures in the
FUV. Column (3): The same as Column (2) in NUV. Column (4):
Magnitudes in the $W1$. Column (5): Magnitudes in the $W2$. }
\end{deluxetable}

\section{Discussion}\label{app}

\subsection{Colors and Comparison with SDSS}
\begin{figure*}
   \centering
   \includegraphics[width=0.85\textwidth]{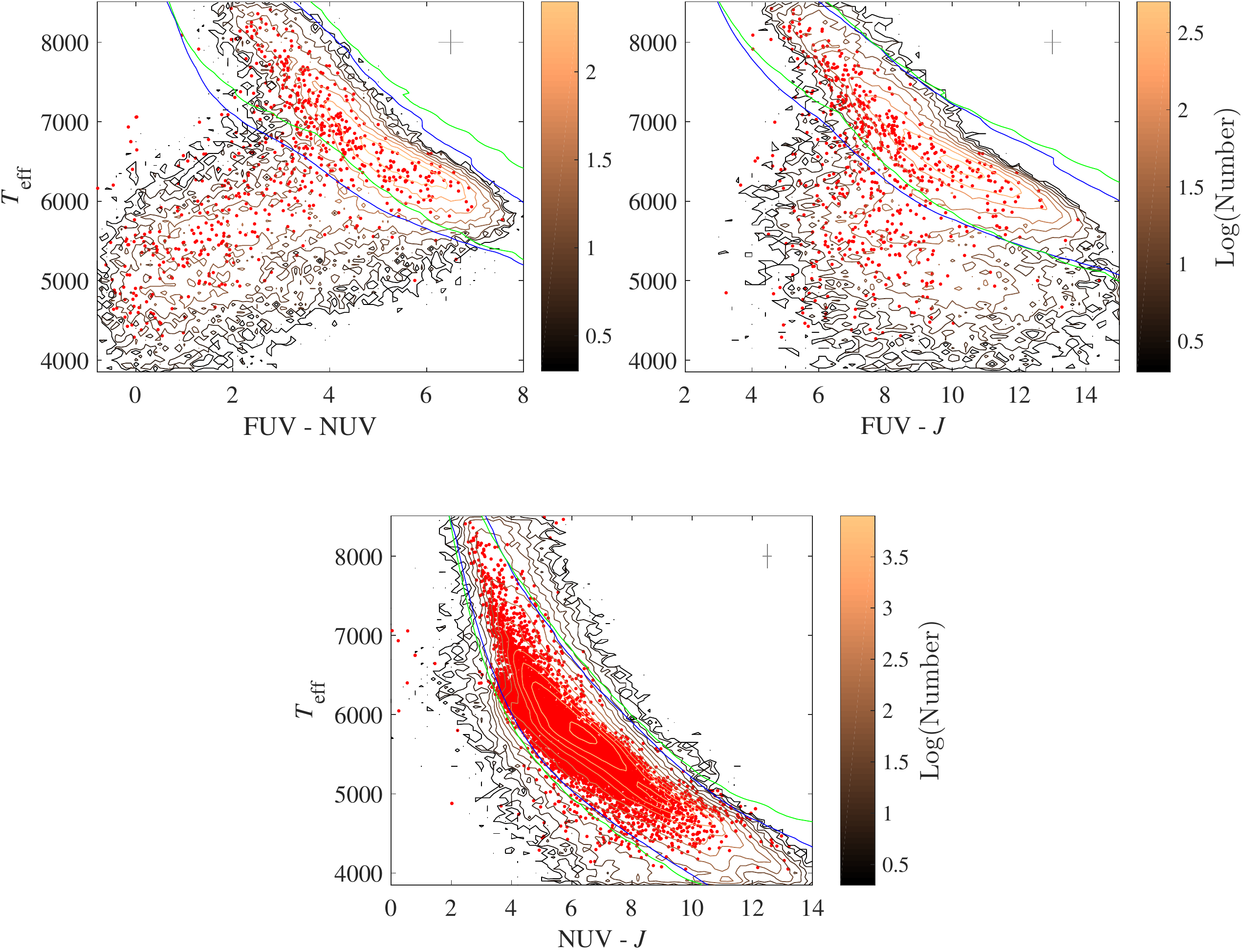}
   \caption{ The effective temperature as functions of the (FUV $-$ NUV), (FUV $-$ $J$)
   and (NUV $-$ $J$) for
   stars in our catalog (contours) and the matched stars processed through the
   SSPP in the SDSS archive (red points).
   The blue solid lines stand for
   the color ranges from PHOENIX photospheric models, and those from ATLAS9 models \citep{Castelli97} are
   shown in green for comparison.
   The average errors are presented as grey crosses in the upper right corners.
   \label{fuvnuv_Teff_SDSS_BTCond}
    }
\end{figure*}

We present color distributions in Figure~\ref{fuvnuv_Teff_SDSS_BTCond}.
The (FUV $-$ NUV) declines with increasing effective temperature for
stars with {\Teff} $>$ 7000 K,
because their UV fluxes are dominated by radiation from stellar photospheres.
This trend disappears and the UV color distribution becomes dispersive for stars with {\Teff} $\sim$ 6000 K,
since their UV colors from upper atmosphere emission probably became comparable to photospheric UV colors.
Some inactive G stars follow the colors predicted by the model, while other active stars have blue excesses.
These active G stars might be either in active close binaries or young single stars \citep{Smith14}.

For stars later than G, the color distributions exhibit remarkably blue excesses,
because of the additional bluer UV emission from the upper regions of atmosphere.
The UV colors of these stars are (FUV $-$ NUV) $\sim$ 1, similar to
the blue peak in \citet{Smith14}, which might imply the general colors of the stellar active regions.

When integrate the distributions of all stars with different types,
we also detect the bimodal distribution of (FUV $-$ NUV) in our
sample, similar to \citet{Smith11}. The bimodal distribution
is perhaps due to the two sub-populations, the UV red stars having
weak activities and mid effective temperatures, and the UV blue
stars having strong activities or high effective temperatures
\citep{Smith14}.

The peaks of the distributions are anti-correlated to effective
temperatures in the figures of (FUV $-$ $J$) and (NUV $-$ $J$), but
we can see obvious blue excesses for the stars with {\Teff}
$\lesssim$ 6000 K and {\Teff} $\lesssim$ 5500 K in the two
figures, respectively. This implies the beginning at which the
radiation from upper regions of the atmosphere dominates
the UV flux. For stars with higher effective temperatures, the UV
fluxes are mainly from stellar photospheres, where the UV-IR colors
depend on the effective temperatures rather than the upper
atmosphere emission.

There are some late A stars redder than the colors predicted by
theoretical models in the distribution of (NUV $-$ $J$). The nature
of the bias is still unclear, which might indicate some
deviation between observation and theoretical templates for the
radiation from stellar photospheres, or amount to potential
spectrally unresolved companion stars with slightly lower effective
temperature than primary stars.

We cross match the stars having valid stellar parameters in our
catalog with the SDSS stars processed through the SEGUE Stellar
Parameter Pipeline (SSPP, \citealt{Lee08a},\citealt{Lee08b}), in
order to check the difference in the effective temperatures
derived from two independent pipelines. In
Figure~\ref{fuvnuv_Teff_SDSS_BTCond}, there is no obvious
systematical deviation in the color distributions. We can conclude
that the distributions of the colors associated with UV are
independent in the stellar parameter pipelines, and
indicate intrinsic properties of our sample. Here we do not
cross identify with APOGEE stars, since they are mainly giants with
effective temperatures in the range from 4000 to 5500 K
\footnote{The stellar parameters from APOGEE exhibit good
consistency with those from the LAMOST \citep{Luo15}. }.

\subsection{Scientific Applications}
Because of its large size and homogeneous nature, this
catalog not only provides a wealth of information
for the study of upper atmosphere emission, but also
enables a time-domain exploration in the UV. In this section, we
would like to discuss some scientific applications of the
catalog.
\subsubsection{A Passible UV Reference Frame}
The MK spectral classification sets the general reference
frame for stars on physical parameters,  and further plays a
critical role in identification of the truly peculiar stars
which do not fit comfortably into the frame \citep{Gray09,Gray14}.
However, the reference frame is poorly constrained in the UV band
due to small samples of stellar emission from upper regions of
atmospheres. Using this large catalog, we could present a
possible UV reference frame based on radiation from all
regions of stellar photospheres.

\begin{figure}
   \centering
   \includegraphics[width=0.5\textwidth]{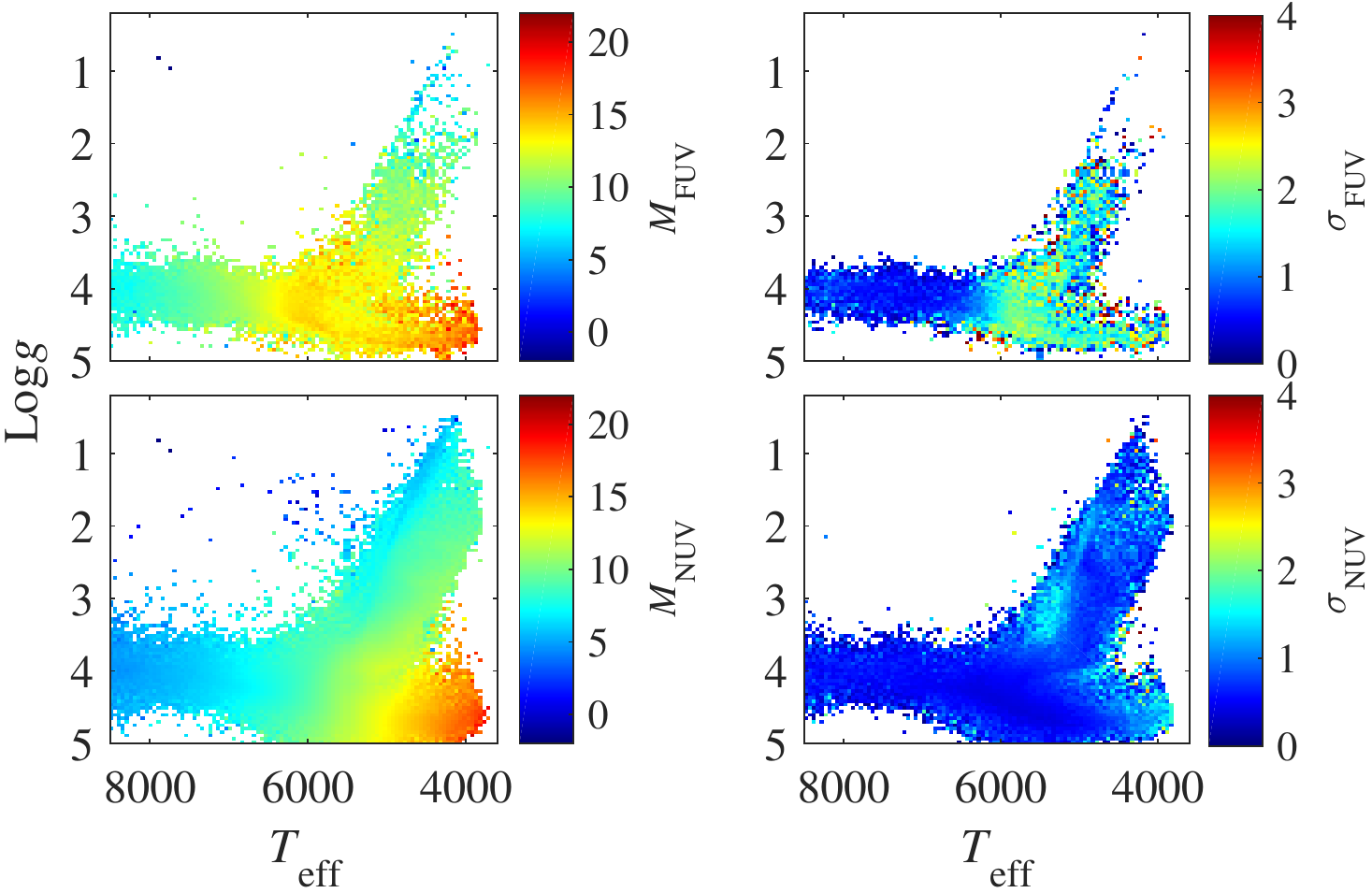}
   \caption{The UV absolute magnitudes in an HR-like diagram.
            Left panels: the distributions of absolute magnitudes in the FUV and NUV.
            Right panels: the standard deviation in the bins of {\Teff} and log $g$.
   \label{HR_M_m}}
\end{figure}

We calculate the absolute magnitudes for stars with valid parallaxes
\citep{Wang16a}, and average the fluxes in each bin of {\Teff} and
log $g$ (left panels of Figure~\ref{HR_M_m}). The UV absolute
magnitude declines with the increase of the effective
temperature, which is similar to the results of \citet{Shkolnik13},
while the standard deviation is larger in the FUV than the NUV for
stars with {\Teff} $\lesssim$ 6000 K.
Such standard deviation is probably due to their strong intrinsic
variability, as shown in the upper two panels in Figure\ref{HR_S_S}.
Because of the small standard deviation and homogeneity in
the stellar parameters, the NUV absolute magnitudes could be used as
a possible reference frame to calibrate the stellar NUV
emission. Such NUV emission includes the stellar radiation not only
from the photosphere but also the upper regions of the atmosphere.

\subsubsection{Colors of Giant and Dwarf Stars}\label{giants}
Giant stars, owing to their high luminosities, are popular tracers
of substructures in the Milky Way, and help us address important
questions, e.g., the size of our Galaxy
\citep{Bochanski14}, the shape of the dark matter halo
\citep{Piffl14}, substructures in the outer Galactic halo
\citep{Sheffield14}, the metallicity distribution \citep{Hayden15},
etc.

\begin{figure}
   \centering
   \includegraphics[width=0.45\textwidth]{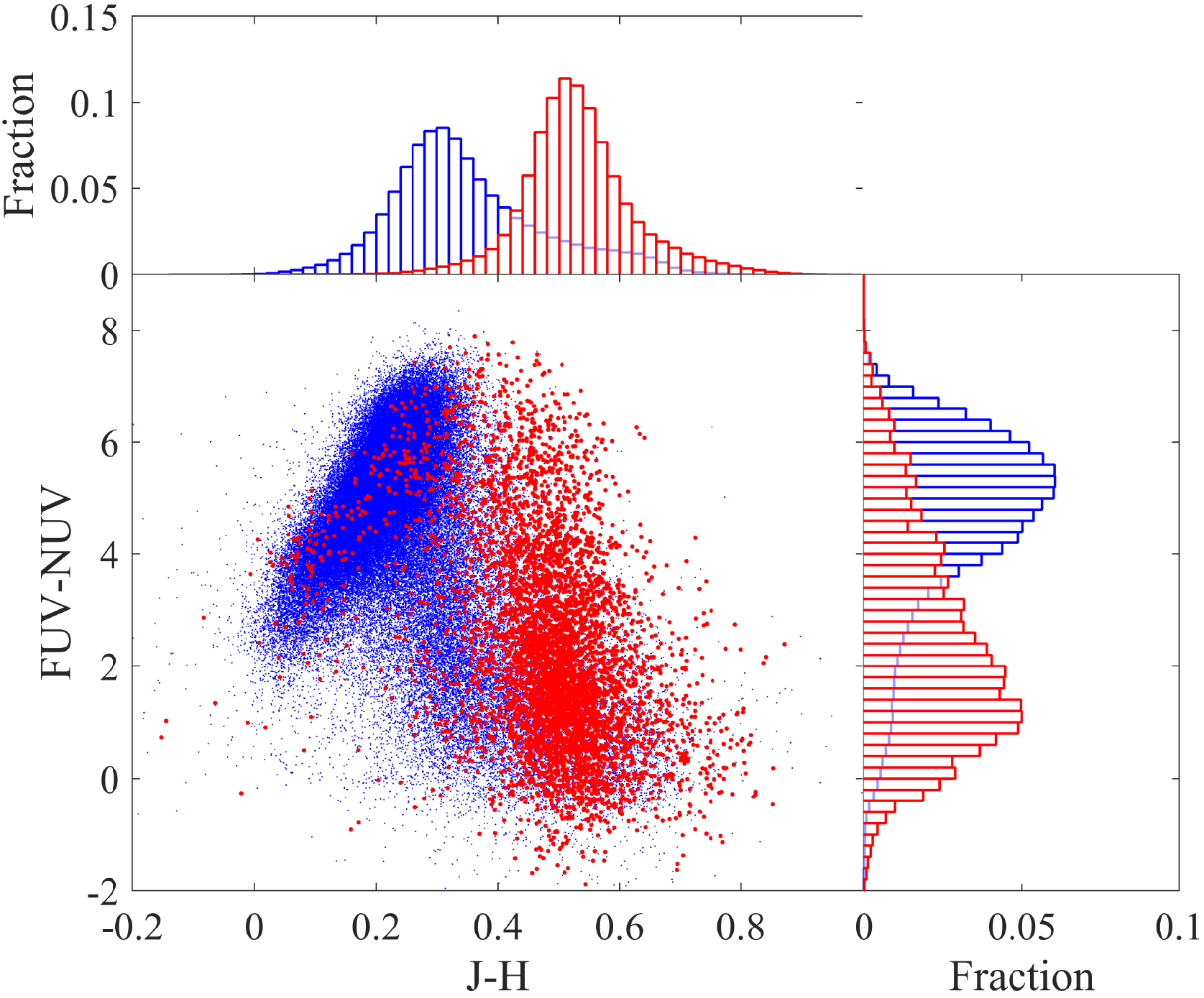}
   \caption{The color-color diagram for the giants (red) and dwarfs (blue) in our sample, and a clear
            $\Lambda$-shape distribution can be seen for dwarfs.
            The upper and right panels show histograms of the colors.
   \label{Color_AFGK_m}}
\end{figure}

We use the criterion provided by \citet{Ciardi11} to separate giants
from dwarfs. Neither the UV involved colors
(Figure~\ref{Color_AFGK_m}) nor the UV excesses can provide
a good separation between giants and dwarfs in our sample. The
distribution of dwarfs exhibits a clear '$\Lambda$' shape. The stars
in the right branch are mainly K dwarfs, which overlap with the
giants. The UV involved colors could separate the stars with {\Teff}
$\sim$ 6000 K, located in the red valley in
Figure~\ref{HR_Color_m}, from other stars with higher temperatures
or stronger upper atmosphere emission. UV and IR photometry
based methods cannot provide a good separation,
but some spectra related methods could distinguish giants,
such as spectral line features \citep{Liu14}.

\begin{figure}
   \centering
   \includegraphics[width=0.45\textwidth]{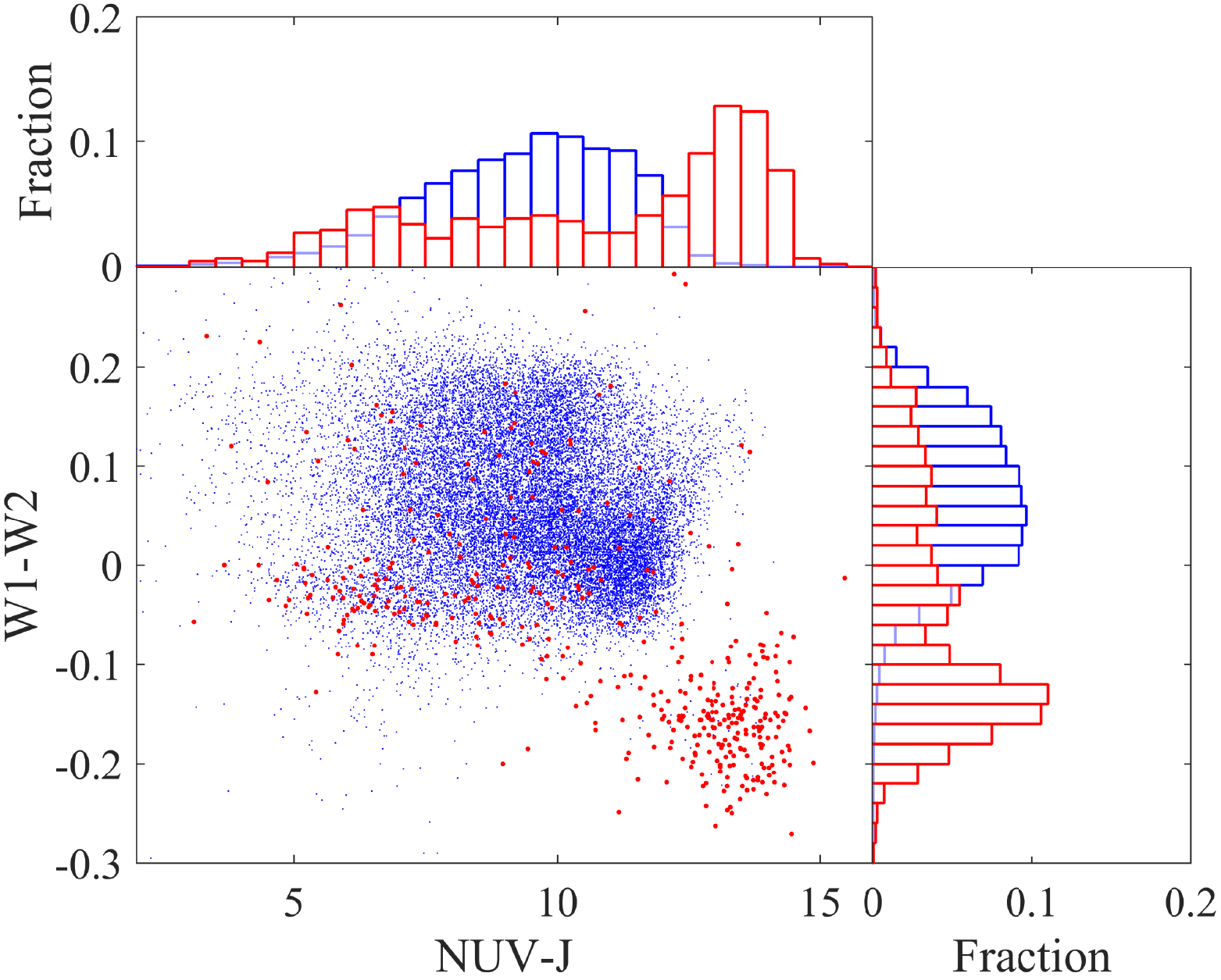}
   \caption{The color-color diagram for the M giants (red) and M dwarfs (blue) in our sample.
            The upper and right panels show the histograms of the colors.
   \label{Color_dM_m}}
\end{figure}

\begin{figure}
   \centering
   \includegraphics[width=0.45\textwidth]{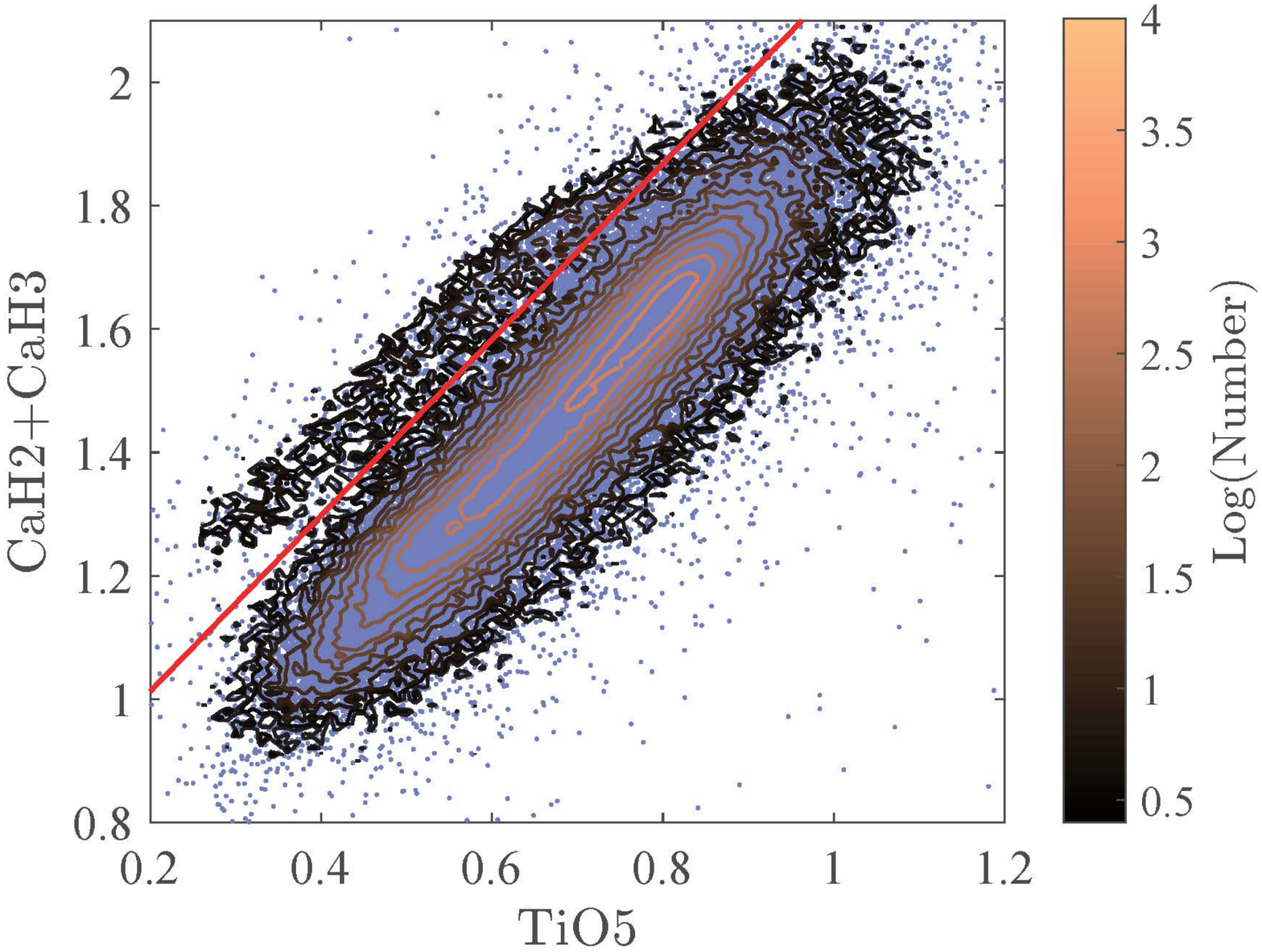}
   \caption{The diagram of the CaH2$+$CaH3 versus TiO5 for M stars in our sample (light blue dots).
            The density contours are the sample in the M stars catalog, and the separation between
            giants and dwarfs is shown as the red line. The M stars above the red line are giants.
   \label{TiOCaH}}
\end{figure}

However, the color of the (NUV $-$ $J$) can separate M
giants from M dwarfs in Figure~\ref{Color_dM_m}. Here we use a
robust linear separation, the lowest density of the stars in
the diagram of molecular-band indices in Figure~\ref{TiOCaH}
\citep{Zhong15}. The upper atmosphere emission of giant
star is weaker than those of dwarfs, and the colors
constructed by both UV and IR bands could provide a good statistical
separation.

\subsubsection{RR Lyrae Stars}
RR Lyr Stars are pulsating periodic horizontal-branch variables with
great variation in the UV in the range of 2$-$5 mag
\citep{Wheatley12}, making them more likely to be detected in the
UV. They are thus popular tracers of Galactic structures, the halo
\citep{Sesar10b} and streams \citep{Drake13}.

\begin{figure}
   \centering
   \includegraphics[width=0.45\textwidth]{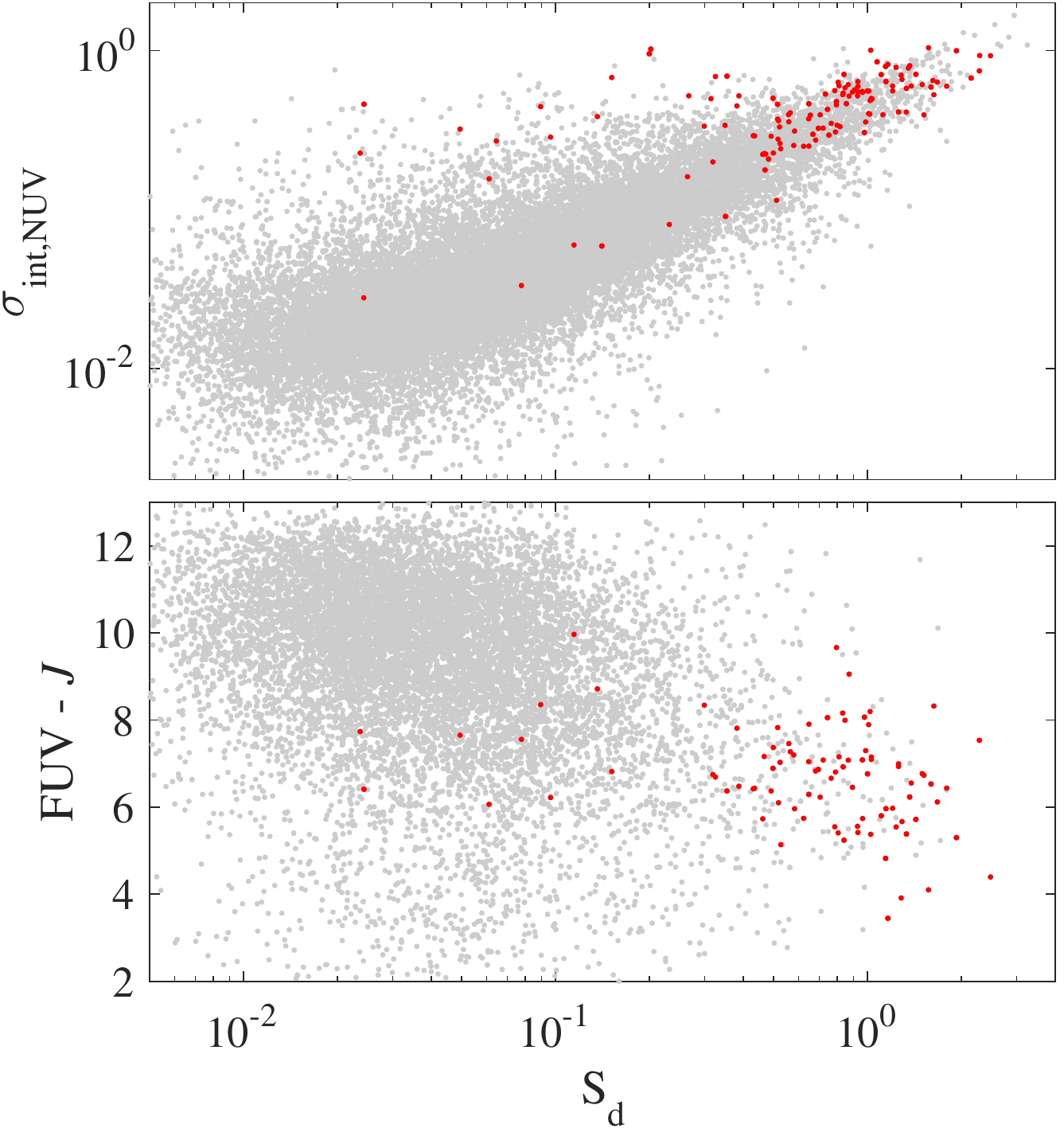}
   \caption{The properties of RR Lyrae stars in our sample.
            Upper panel: the diagram of $\sigma_{\rm int}$ versus $S_{\rm d}$ of A
            and F stars in our sample (the gray dots), and
            the matched RR Lyrae stars of \citet{Drake13} and \citet{Abbas14} (the red dots).
            Lower panel: the (FUV $-$ $J$) as a function of the $S_{\rm d}$.
   \label{RR}}
\end{figure}

\begin{figure*}
   \centering
   \includegraphics[width=0.9\textwidth]{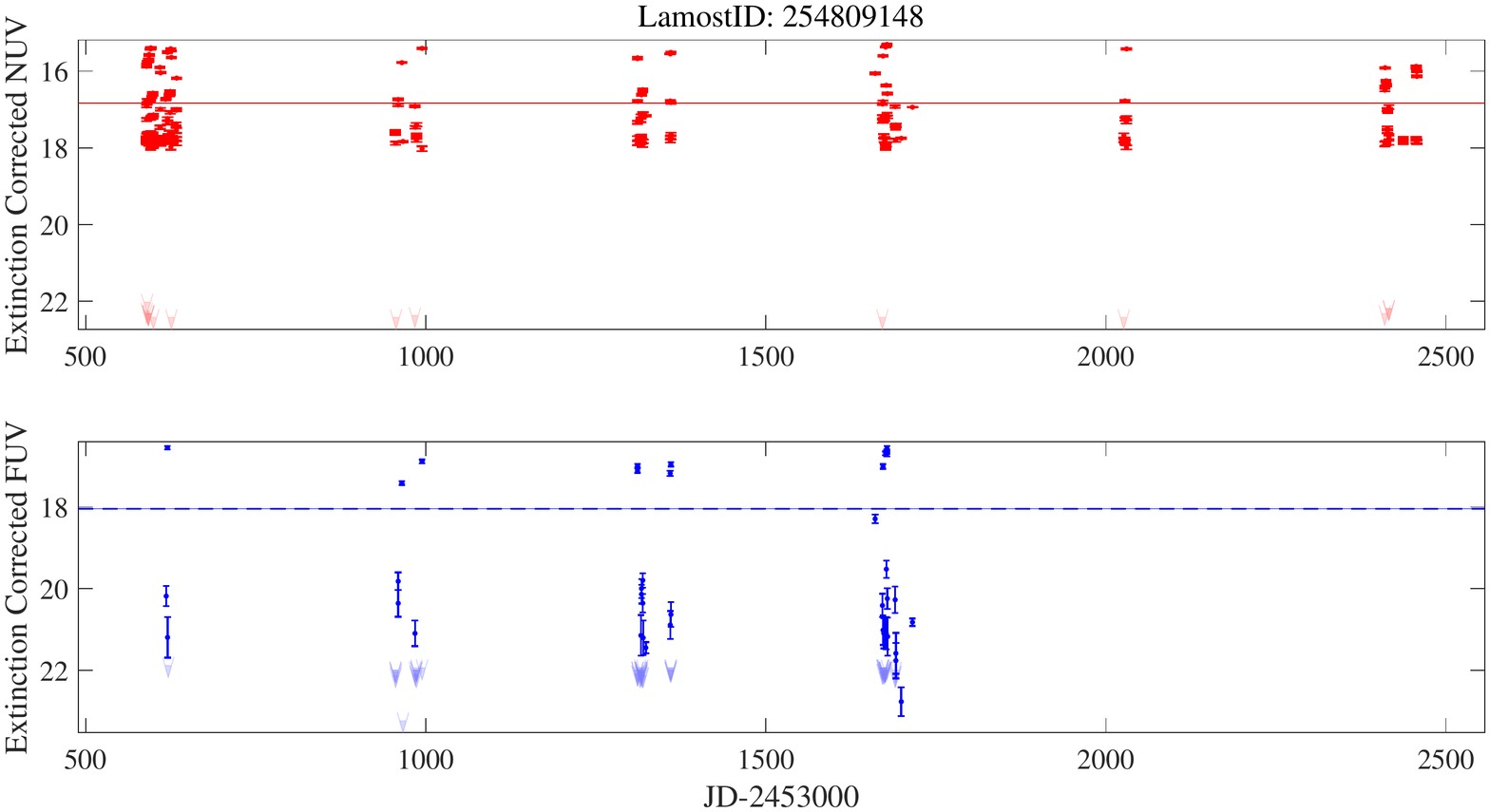}
   \caption{The light curves of an RR Lyrae star in our sample in the FUV (blue) and NUV (red). The error bars are
            magnitudes in the visit exposures, and the arrows are undetected upper limit magnitude in the visit exposures.
            The solid and hashed lines are mean magnitudes and their errors respectively.
   \label{LC254809148}}
\end{figure*}

We cross match the RR Lyr stars in \citet{Drake13} and
\citet{Abbas14} with our sample, and plot them over the A and F
stars in Figure~\ref{RR}. The UV counterparts of RR Lyr stars are
concentrated in the region of high $S_{\rm d}$ and $\sigma_{\rm{int,
NUV}}$, and have a bluer color of (FUV $-$ $J$) than other A and F
stars. In Figure~\ref{LC254809148}, we plot the light curve of
an RR Lyr star with LID $=$ 254809148 as an example. The
amplitude of the UV light curve is higher than that in the
optical band by a factor of 2 \citep{Sesar10a}.

\section{Summary and Further Work}\label{sum}
We provide a catalog of over three millon $GALEX$-observed stars,
and the parameters to quantify stellar activity and variability,
based on well estimated background and extinction, a newly developed
method on upper limit magnitudes, spectral properties from
the LAMOST, and time-resolved photometry in the FUV and NUV. This
includes 2,202,116 detected and 889,235 undetected stars. The
occurrence of possible false positives is below 1.3\% in our sample,
and the candidates of spectrally unresolved WDMS are selected from
the UV involved color-color diagram.

The emission from regions beyond the stellar photosphere
could result in the discrepancy between observed UV minus IR colors
and those predicted by theoretical photospheric models. We quantify
such discrepancy by the UV and normalized UV excesses, and find that
they both decline with increasing effective temperatures, and the
late-type dwarfs tend to have high excesses. Similar to other
studies, the normalized UV excess suffers from large systematic
uncertainties in the photosphere contribution, especially
for stars with high effective temperatures. The systematic
uncertainties of these stars may be from errors in the extinction
and the model interpolation. We present a linear relation
between the (FUV $-$ NUV) and the {\Rfuv}. The UV color as a proxy
of the FUV excess is recommended when the stellar parameter is
unavailable.
The distributions of the parameterized variation in the HR-like diagram shows that the effective
temperature declines with increasing variability.

We also provide the absolute magnitudes in an HR-like
diagram, which could serve as a possible reference frame in the NUV.
With the UV involved colors, the M giants and M dwarfs could be well
separated, but other giants and dwarfs cannot be distinguished with
colors or UV excesses. We find that the RR Lyrae stars, strong UV
emitters, have stronger variation than most A and F stars in our
sample.

The stellar parameters are available for a portion of late A, F, G
and K stars in our catalog, which are applied to estimate the UV and
normalized UV excesses. In order to investigate the excesses for
stars with other spectral types, we could introduce other catalogs,
e.g., the further data released of the LAMOST Spectroscopic Survey
of the Galactic Anticentre (LSS-GAC, \citealt{Yuan15}) in which the
parameters are derived from LSP3 \citep{Xiang15}, and a more
complete version of M stars from LAMOST \citep{Zhong15} in which a
template matching technique is used. The UV emission of B, A and M
stars will be characterized in detail in our further studies.

We are also going to study the stars located in the grey circle in Figure~\ref{HR_Color_m}, with time-resolved
spectra in order to check their binarity. The further mid-resolution spectra in LAMOST survey may shed light
on the rotations of these stars.


\begin{acknowledgements}
We are grateful to Krzysztof Findeisen, Beate Stelzer and Murthy Jayant for valuable discussions.
This work was supported by the National Program on Key Research and Development
Project (Grant No. 2016YFA0400804),
the National Natural Science Foundation of China (NSFC)
through grants NSFC-11603038/11333004/11425313/11403056, and the National Astronomical Observatories,
Chinese Academy of Sciences (NAOC) under the Young Researcher Grant.
Y.W. acknowledges the fund supplied by the Guangdong Provincial Engineering Technology Research Center for Data Science.
Some of the data presented in this paper were obtained from the Mikulski Archive for
Space Telescopes (MAST). STScI is operated by the Association of Universities for Research
in Astronomy, Inc., under NASA contract NAS5-26555. Support for MAST for non-HST data is provided
by the NASA Office of Space Science via grant NNX09AF08G and by other grants and contracts.

The Guoshoujing Telescope (the Large Sky Area Multi-Object
Fiber Spectroscopic Telescope, LAMOST) is a National Major
Scientific Project which is built by the Chinese Academy of
Sciences, funded by the National Development and Reform Commission,
and operated and managed by the National Astronomical Observatories,
Chinese Academy of Sciences.

Funding for SDSS-III has been provided by the Alfred P.
Sloan Foundation, the Participating Institutions, the National
Science Foundation, and the U.S. Department of Energy Office
of Science. The SDSS-III Web site is \url{http://www.sdss3.org/}.

SDSS-III is managed by the Astrophysical Research
Consortium for the Participating Institutions of the SDSS-III
Collaboration including the University of Arizona, the
Brazilian Participation Group, Brookhaven National Laboratory,
University of Cambridge, Carnegie Mellon University,
University of Florida, the French Participation Group, the
German Participation Group, Harvard University, the Instituto
de Astrofisica de Canarias, the Michigan State/Notre Dame/
JINA Participation Group, Johns Hopkins University, Lawrence
Berkeley National Laboratory, Max Planck Institute for
Astrophysics, Max Planck Institute for Extraterrestrial Physics,
New Mexico State University, New York University, Ohio
State University, Pennsylvania State University, University of
Portsmouth, Princeton University, the Spanish Participation
Group, University of Tokyo, University of Utah, Vanderbilt
University, University of Virginia, University of Washington,
and Yale University.
\end{acknowledgements}


\end{document}